%% file: main_revision.tex
\documentclass[sigconf]{acmart}
\usepackage{enumitem}
\newcommand{\name}{Vista}
\usepackage{tikz}
\usepackage{subcaption}
\usepackage[ruled,vlined]{algorithm2e}

\DontPrintSemicolon
\usepackage{amsmath}
\DeclareMathOperator*{\argmax}{arg\,max}

\usepackage{enumitem}

\newcommand{\newchanges}[1]{\textcolor{black}{#1}}

\definecolor{dark_green}{RGB}{0, 120, 10}
\newcommand{\sssec}[1]{\vspace*{0.05in}\noindent\textbf{#1}}

\renewcommand\footnotetextcopyrightpermission[1]{} % removes footnote with conference information in first column
\begin{document}
\sloppy
\title{Low-latency Imaging and Inference from LoRa-enabled CubeSats}
% Sensing Product Integrity without Opening the Box using Non-Invasive Acoustic Vibrometry
\author{Akshay Gadre}
 \affiliation{\institution{Carnegie Mellon University}
 \country{USA}}
 \email{agadre@andrew.cmu.edu}
 
\author{Swarun Kumar}
 \affiliation{\institution{Carnegie Mellon University}
 \country{USA}}
 \email{swarun@cmu.edu}
 
 \author{Zachary Manchester}
 \affiliation{\institution{Carnegie Mellon University}
 \country{USA}}
 \email{zacm@cmu.edu}

% \author{Deepak Vasisht}
% \affiliation{\institution{Microsoft and UIUC}}
% \email{deepakv@illinois.edu}

% \author{Nikunj Raghuvanshi}
% \affiliation{\institution{Microsoft}}
% \email{nikunjr@microsoft.com}

% \author{Bodhi Priyantha}
% \affiliation{\institution{Microsoft}}
% \email{bodhip@microsoft.com}

% \author{Manikanta Kotaru}
% \affiliation{\institution{Microsoft}}
% \email{mkotaru@microsoft.com}

% \author{Swarun Kumar}
% \affiliation{\institution{Carnegie Mellon University}}
% \email{swarun@cmu.edu}

% \author{Ranveer Chandra}
% \affiliation{\institution{Microsoft}}
% \email{ranveer@microsoft.com}

% \keywords{Acoustic Vibrometry, Product Integrity Testing, Non-Invasive Acoustic Sensing}

% \copyrightyear{2021}
% \acmYear{2021}
% % \setcopyright{rightsretained}
% \acmConference[MobiCom '21]{The 27th Annual International Conference on Mobile Computing and Networking}{October 25--29, 2021}{New Orleans, LA}
% \acmBooktitle{The 27th Annual International Conference on Mobile Computing and Networking (MobiCom '21), October 25--29, 2021, New Orleans, LA}\acmDOI{XX.XXXX/XXXXXXXX.XXXXXXX}
% \acmISBN{XXX-X-XXXX-XXXX-X/XX/XX}

%\renewcommand{\shortauthors}{}

\begin{abstract}
 %\footnote{\underline{Mi}crosoft \underline{L}ook \underline{T}hrough and \underline{On}ward}
Recent years have seen the rapid deployment of low-cost CubeSats in low-Earth orbit, primarily for research, education, and Earth observation. The vast majority of these CubeSats experience significant latency (several hours) from the time an image is captured to the time it is available on the ground. This is primarily due to the limited availability of dedicated satellite ground stations that tend to be bulky to deploy and expensive to rent. This paper explores using LoRa radios in the ISM band for low-latency downlink communication from CubeSats, primarily due to the availability of extensive ground LoRa infrastructure and minimal interference to terrestrial communication. However, the limited bandwidth of LoRa precludes rich satellite Earth images to be sent -- instead, the CubeSats can at best send short messages ($\sim$ a few hundred bytes).

% A significant number of these, including launches by NASA, have used LoRa radios that allows for low-latency communication of short messages with the ground, primarily due to the availability of extensive ground LoRa infrastructure and minimal interference to terrestrial communication. However, LoRa's communication bandwidth and reliability is extremely poor due to the long range and high Doppler shifts in the CubeSat context. This precludes some of the most interesting applications for CubeSat communication, including satellite image delivery and processing. 

This paper details our experience in \newchanges{communicating with a LoRa-enabled CubeSat launched by our team}. We present \name, a communication system that makes software modifications to LoRa encoding onboard a CubeSat and decoding on commercial LoRa ground stations to allow for satellite imagery to be communicated, as well as wide-ranging machine learning inference on these images. This is achieved through a LoRa-channel-aware image encoding that is informed by the structure of satellite images, the tasks performed on it, as well as the Doppler variation of satellite signals. A detailed evaluation of \name\ that builds upon wireless channel measurements from \newchanges{the LoRa-CubeSat launch (in 2021)} shows 4.56 dB improvement in LoRa image PSNR and 1.38$\times$ improvement in land-use classification over those images.
\end{abstract}
\maketitle
\pagestyle{empty}
\input{text/1-introduction}

\input{text/4-overview}

\input{text/6-loragroundstation}
\input{text/5-imageprocessing}
\input{text/7-limitationsanddiscussions}

\input{text/8-implementation}

\input{text/9-results}

\input{text/2-relatedwork}

\input{text/10-conclusion}
\newpage

% \newpage
% \vspace*{-0.1in}
\bibliographystyle{ACM-Reference-Format}
\bibliography{myref}

\end{document}

%% file: text/1-introduction.tex
\section{Introduction}
%  \begin{center}
%  \vspace*{0.1in}\textit{ I have taken a moment here to rest, \\ to steal a view of the glorious vista that surrounds me,\\ to look back on the distance I have come.\\ \vspace*{0.03in} --  Nelson Mandela}
%  \end{center}

The massive reduction in costs of launching satellite payloads~\cite{cheaplaunch} has led to a proliferation of CubeSats in Low Earth Orbit and truly democratized access to space for small players -- researchers, students and satellite enthusiasts. Today, such CubeSats target varied geo-sensing and imaging applications and number in the thousands\cite{alotofsats}. However, a key question that bottlenecks for CubeSats from small players is the answer to a simple question: ``Is my satellite online right now, and if so, what is it imaging?''. Most current CubeSats rely on existing wireless technologies, developed for larger satellites, for downlink communication. However, ground stations for these technologies are extremely expensive to rent/install for small teams -- with most teams able to downlink from CubeSats quite rarely -- often once per few hours~\cite{vasisht2021l2d2}, i.e. whenever a ground station is in range. Imagine a time-critical event that occurs during this long window where the satellite is simply inaccessible. For instance, consider an interesting Earth image that needs to be sent downlink. Such images could be time-sensitive, capturing on-Earth emergencies such as flooding or weather events around specific locations on the ground observed by the CubeSat.  In this paper, we study if low-latency access to satellites can be enabled by \textit{re-using} existing, global ground radio infrastructure rather than building one from scratch -- specifically, commercial LoRa base stations. In particular, we see a LoRa chip on CubeSats as a \textit{quick and dirty} add-on to push timely information from a CubeSat to the ground anywhere in the world -- \textit{not} as a full-on replacement satellite communication system.

\begin{figure}[t]
    \centering
    \includegraphics[width=\linewidth]{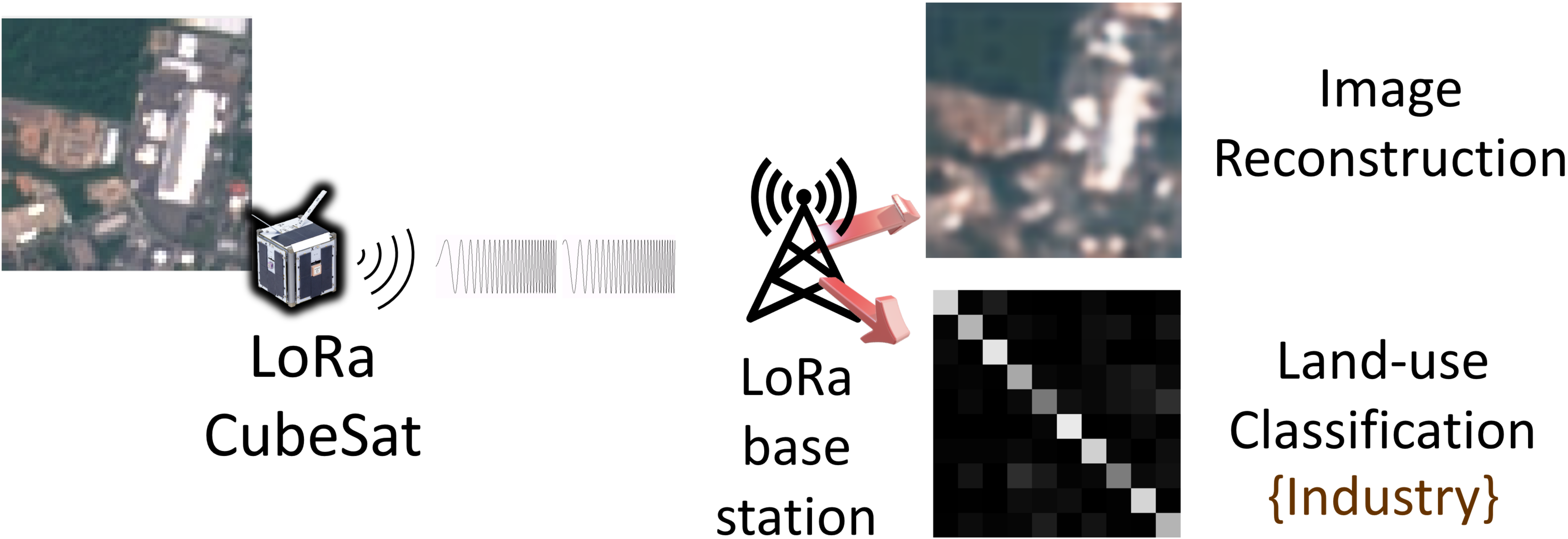}
    \vspace*{-0.25in}
    \caption{\name\ enables practical imaging and inference applications for LoRa CubeSats within a single packet}
    \label{fig:dopplervariance}
    \vspace*{-0.15in}
\end{figure}
%In fact, CubeSats have even been launched to Mars\cite{} and are expected to play a big role in the upcoming Artemis missions\cite{}. These CubeSats have been used to transmit imagery, act as signal relays and amplifiers, and create a mesh of interconnected base stations for narrowband LP-WAN technologies. MarCO WALL-E CubeSat, in fact, became one of the first CubeSats to communicate images (at Mars) in real-time via relays using a massive X-band (8-12 GHz) antenna\cite{}. Unfortunately, integration of imaging on CubeSats in low-earth-orbit (LEO) has been comparatively much slower. A key factor behind this is the fact that many of these LEO CubeSats, typically launched by hobbyists, are much smaller and use off-the-shelf hardware, reducing their communication capabilities with the aim of keeping costs low. A popular off-the-shelf technology being touted and used by most hobbyists in the 915MHz and 2.4 GHz ISM band is LoRa.

The reasons behind why LoRa is an attractive wireless technology to communicate with CubeSats in a timely manner is three-fold: (i) There is a vast existing ground infrastructure that continuously listens for LoRa packets that can be used to receive packets around the globe. (ii) It's chirp-spread-spectrum narrowband modulation is immune to much of the on-ground communication in these bands. Further, it has been shown that other technologies also are affected minimally due to LoRa traffic at these frequencies \cite{voigt2016mitigating,chen2021lofi}. (iii) It accommodates a large number of transmitters simultaneously. 

Yet, there are two key challenges in communicating data from LoRa-equipped CubeSats to existing ground stations -- the limited bandwidth of LoRa and a highly lossy wireless medium. These stem from a mismatch between the typical LoRa client (a mostly static terrestrial device deployed a few kilometers away) and a CubeSat (a device in space moving at seven kilometers per second and hundreds or thousands of kilometers away). In fact, theoretical analysis of the impact of these Doppler effects on LoRa at high frequencies\cite{doroshkin2019experimental} has shown high amount of packet drops and error in detection. Indeed, that is why most of the recently launched LoRa CubeSats~\cite{swarm,fossat1,gossamer,sattala} operate at lower frequencies (137 MHz and 407-450 MHz) and leverage specialized receivers instead of leveraging existing 915 MHz ISM band ground infrastructure.

In this paper, we describe our experience in building wireless communication link from LoRa-enabled CubeSats in low-earth orbit designed to better communicate with existing ground LoRa infrastructure in the ISM band. We present \name, a CubeSat communication system that uses conventional LoRa modulation to directly deliver image information with high efficacy, by overcoming the challenges of limited bandwidth and infrastructure compatibility. We first study the performance of LoRa communication in the ISM band through the development and launch of a LoRa CubeSat. We then use wireless channel measurements from this CubeSat in space to then demonstrate that \name\ can provide an effective 55.55\% lower latency, 4.56 dB improvement in image retrieval SNR and 1.38$\times$ improvement in classification tasks over the images.

The rest of this paper details \name's approach to address these challenges, namely: (1) Enabling existing LoRa ground station infrastructure to detect and decode CubeSat LoRa signals. (2) Designing channel-aware solutions for encoding images and performing inference on these images over the bandwidth-starved CubeSat LoRa links.

\vspace{0.05in}\noindent\textbf{Connecting LoRa CubeSats with ground infrastructure:} Engineering a terrestrial LP-WAN technology to communicate at a range fifty times larger than its conventional range is a challenging task. We require a precise optimization of the design space at the satellite to maximize the throughput of the link by choosing the correct configuration in terms of bandwidth and spreading factor. Further, we must remain compatible with existing ground stations to effectively detect and decode these packets. At the CubeSat, we describe how a LoRa CubeSat should be designed to overcome challenges with motion in space -- specifically as small satellites of about 1U (1000 cubic centimeters) tumble quite significantly in orbit.  We show how these constraints inform antenna design as well as incur spectrum policy considerations on the ground as well as space. At the ground stations, \name\ addresses the problem of the large attenuation and massive Doppler shift that render the packets from CubeSats undecodable at pre-existing LoRa base stations.  \name\ mitigates this by leveraging existing hardware modules that operate at various different bandwidths -- that LoRa traditionally provides as part of its data-rate adaptation pipeline. We show how a careful selection of these pre-existing correlation modules can enable terrestrial ground stations to detect these missed packets with a simple software update. Sec.~\ref{sec:dop} describes these solutions for both satellite ground stations and the LoRa CubeSat. % %The rest of Sec.~\ref{sec:dop} describes other such software-only ground station and cubesat modifications that provide resilience to interference and various sources of noise.
%solutions that are resilient to the above challenges to even stand a chance of communicating imagery downstream.

\vspace{0.05in}\noindent\textbf{Channel-Aware Image Encoding:} Given this bandwidth-starved CubeSat link, it is critical to maximize the utility of every byte to communicate data downlink. We specifically focus on communicating image data for the purpose of performing inference tasks. A n\"aive solution to compress images captured from the camera would be to use traditional compression algorithms such as JPEG with appropriate compression ratio. Yet, there exists an exciting opportunity in the satellite image context, where rich prior information is available in how these images are structured, e.g. image data from NOAA LEO weather satellites~\cite{noaa}. Further, compression of image data can also be actively guided by the nature of machine-learning tasks performed on these images. Most importantly, one can also account for the nature of the wireless channel between CubeSats and the LoRa ground stations, where factors such as intra-packet Doppler shift, noise, and satellite trajectories play a key role in the nature of image bit errors one can expect. In other words, rather than building a channel, data and task-agnostic image compression scheme, \name\ designs an image encoding that is informed by the general nature of wireless channels, the structure of satellite imagery, and the tasks performed on this data. 

Unlike prior work in joint source-channel coding for wireless image transmission\cite{guionnet2003soft,chandramouli1998adaptive,burlina1998error,debrunner2000error,bourtsoulatze2019deep} which deals with general uniform error distributions, operating LoRa (a technology that encodes data in frequency shifts) in 915 MHz ISM band (large amount of Doppler) leads to a non-uniform distribution of errors across bits.  
Sec.~\ref{sec:chaware} describes our design of a data-task-driven autoencoder which is optimized for such space-to-ground wireless channels. We train this autoencoder with real satellite image data, our satellite trajectory, as well as our own LoRa channel measurements. We further make this encoding resilient to information loss due to discretization and wireless impairments. We provide various design points in our autoencoder that balances the diversity of tasks that can be performed with the image encoding on the ground and the amount of data that needs to be communicated.

We study \name\ through the development of a communication stack and launch of a LoRa-enabled satellite in the ISM band in 2021 \cite{vr3x,zacpaper}. We study and report the overall signal-to-noise ratio, observed Doppler effects and improved ground-station latency of this LoRa satellite. We then collect long-term wireless channel traces measured from the satellite in orbit across an year. We implement, evaluate and emulate various flavors of \name\ on these channel measurements to showcase the potential benefits in terms of latency, packet detection and inference accuracy, demonstrating:
\begin{itemize}
    \item 55.55\% reduction in latency with a coverage across 44.92\% of the CubeSat orbit vs. 33.27\% of existing infrastructure.
    \item Up to 15 dB improvement in SNR for packet detection in high Doppler scenarios.
    \item 4.56 dB PSNR improvement in images retrieved and 38\% improvement in accuracy of ML tasks over images communicated using \name.
\end{itemize}

\vspace{0.05in}\noindent\textbf{Contributions:} Our contributions include: 
\begin{itemize}
\item A novel CubeSat wireless communication system that uses conventional LoRa modulation to enable high-quality image inference over bandwidth starved links.
\item A system that makes software modifications to commercial LoRa ground gateways to detect highly Doppler shifted packets from LoRa CubeSats.
\item Detailed implementation and evaluation through wireless channel measurements from a commercial LoRa enabled satellite launched by our team. 
\end{itemize}

%% file: text/4-overview.tex
\begin{figure*}[!t]
    \centering
    \includegraphics[width=1\linewidth]{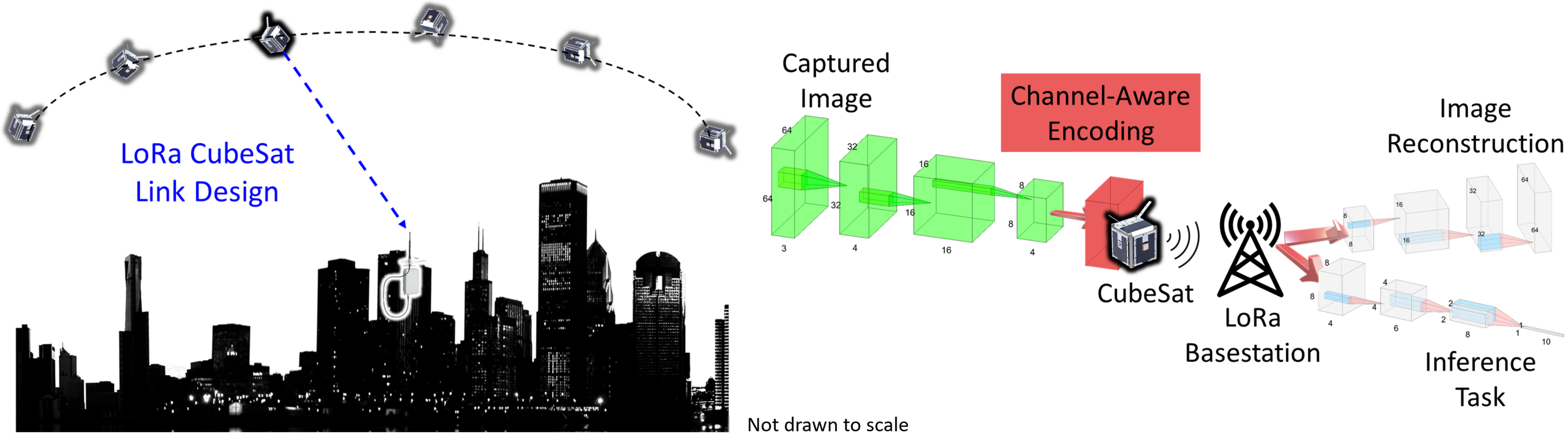}
    
    \caption{Vista develops channel aware data-task driven encodings and Doppler-resilient packet detection to enable fine-grained imaging and low-latency inference from LoRa-enabled CubeSats}
    \label{fig:systemdiagram}
\end{figure*}

\section{Overview}
\name\ is a CubeSat communication system that uses LoRa modulation at the satellite and conventional ground infrastructure, making software-only modifications at both ends to improve delivery and processing of images. Fig.~\ref{fig:systemdiagram} depicts \name's contributions. The rest of the paper discusses the two key components of \name's design. 

\vspace*{0.05in}\noindent\textbf{Enabling LoRa CubeSat -- Ground Station Links: } We showcase the design decisions behind \newchanges{LoRa CubeSats~\cite{zacpaper} built by our team} to communicate in the 915 MHz ISM band. We next show how we can redesign existing commercial off-the-shelf LoRa base stations to detect and decode CubeSat packets accounting for the extremely high Doppler shifts these packets experience. Sec. \ref{sec:dop} describes our solution. %We present  software-only modifications to existing LoRa ground-infrastructure that supports detection of packets that would otherwise be dropped. 

\vspace*{0.05in}\noindent\textbf{Channel-Aware CubeSat Image Encoding: } At the satellite, \name\ encodes satellite images ensuring resilience to Doppler and other wireless impairments, while remaining cognizant of satellite image structure and higher-layer inference desired on images. At the ground stations, \name\ decodes images and/or performs desired inference on the images, such as identifying land use patterns. Sec. \ref{sec:chaware} describes \name's solutions at both the CubeSat and the ground infrastructure.

%% file: text/6-loragroundstation.tex
\section{LoRa CubeSat and Ground Station design}\label{sec:dop}
This section describes the principles behind the LoRa transmitter design on a CubeSat and the challenges and issues to overcome for enabling a robust downlink from the satellite. Further, we describe our approach to empower the ground infrastructure to improve packet reception from CubeSats.

\subsection{LoRa CubeSat Design}\label{sec:loracubesatdesign}

In this section (and this paper), our focus is solely on the LoRa wireless communication related design decisions for the CubeSat. We refer the reader to \cite{zacpaper}\footnote{Reference anonymized for double blind review.  \cite{zacpaper} describes the detailed compute/storage/control system design of the satellite hardware, in contrasts to the LoRa communication aspects which are the sole focus of this paper (\name). } for comprehensive description of other (non-LoRa communication) aspects of the CubeSat design -- including on on-board compute, storage, energy efficiency, etc. We primarily describe some of the key design decisions to keep in mind while building LoRa-enabled CubeSats. At first glance, one would think that simply using existing off-the-shelf LoRa components connected to a power amplifier in the transmission chain would work as a CubeSat transmitter natively. However, there are multiple challenges and design decisions that are critical to optimize the communication downlink from the low-earth orbit (LEO): (1) LEO Spectrum Policy in the ISM band; (2) Antenna Design for the CubeSat transmitter; (3) Choosing Transmission Parameters.

\noindent\textbf{Spectrum Policy:} There are two entities that govern the spectrum in the low-earth orbit. The first is the Federal Communications Commission (FCC)~\cite{fcc} that provides licenses to US-flagged spacecraft (i.e. any US company, university) for communication from satellites. The other entity is National Telecommunications and Information Administration (NTIA)~\cite{ntia} which provides special permissions for experiments on spacecraft backed by US government (i.e. NASA). The only policy-wise complicating factor in 915~MHz band~\cite{fccspectrum} is that, it is specially allocated for federal civil and naval operators for radio-location services among others. However, recent meetings between these two entities point towards building a unified spectrum policy in the near future~\cite{ntiafcc}. The CubeSat built by our team operates in the ISM band for evaluating 915 MHz LoRa downlink communication at a bandwidth of less than 125 KHz.

\begin{figure}[!t]
    \centering
    \vspace*{-0.3in}
    \includegraphics[width=\linewidth]{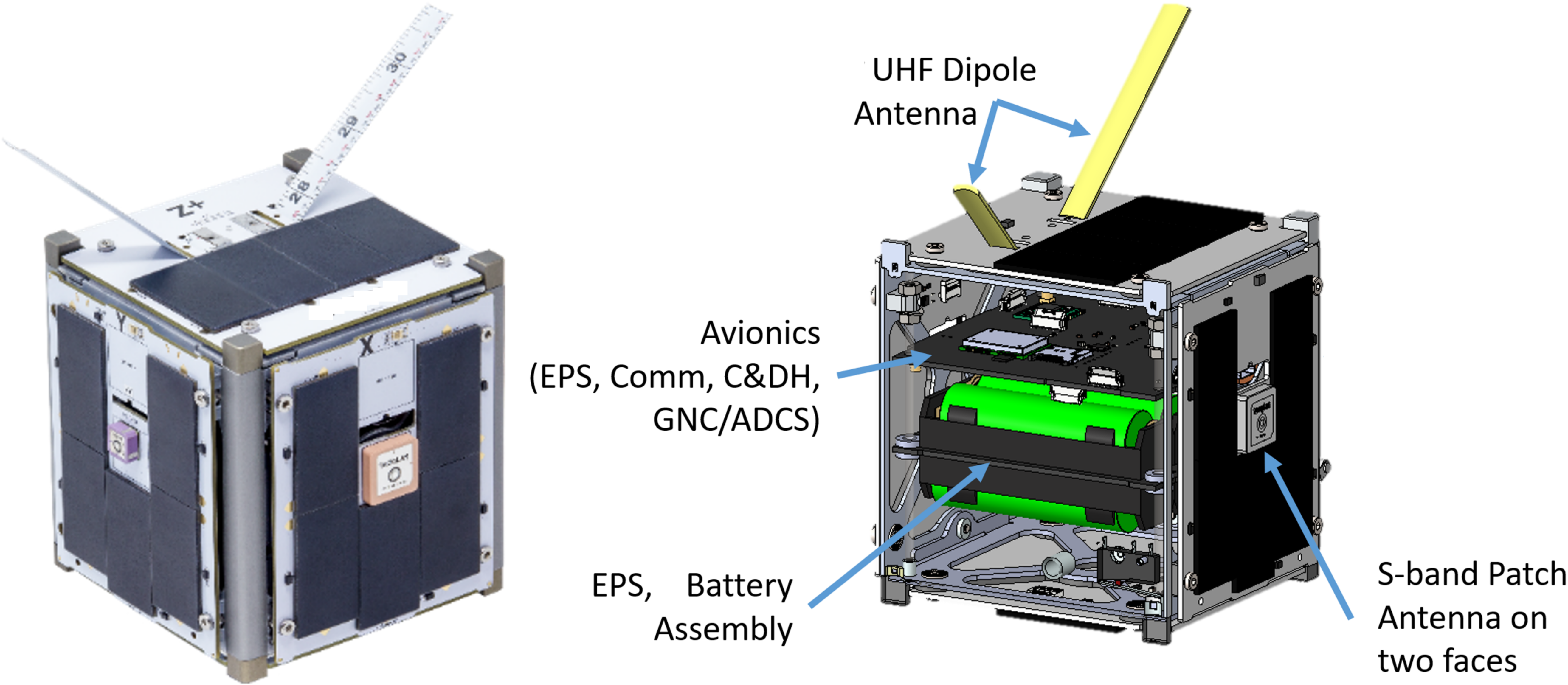}
    \vspace*{-0.2in}
    \caption{\name\ Satellite Design}
    \label{fig:satlabels}
   
\end{figure}

%While the 915 MHz ISM band is typically free to transmit with a few pre-determined constraints for terrestrial transmitters via the Federal Communications Commission (FCC), the policy for aeronautical mobile devices is governed by the National Telecommunications and Information Administration (NTIA). In this context, the 915~MHz band~[] is specially allocated for federal civil and naval operators for radio-location services among others. However, there have been recent draft policy changes in this band towards enabling LoRa based downlinks []. Indeed, this is the reason why existing IoT from Space deployments by companies like Swarm, operate at different lower frequencies. Our cubesat was launched with specific spectrum policy exceptions from NASA to transmit in this band for evaluating the feasibility of 915 MHz LoRa downlink communication. Our specific permissions allowed for global operation at 915.6 MHz and a bandwidth of less than 125 KHz. 

Another obvious concern is to design geo-location awareness. Off-the-shelf GPS typically is locked by CoCom\footnote{Coordinating Committee for Multilateral Export Controls} at altitudes higher than 18 kms and for clients faster than 0.5 m/s. Thus, our satellite used a GPS receiver with CoCom limits removed to ensure compliance. However, the GPS did not lock due to a very important reason  -- small 1 unit CubeSats tumble significantly in space (sometimes up to 7.5 revolutions per minute).  Thus, providing  geo-location awareness for spectrum usage to such small satellites remains an open challenge. Our LoRa CubeSat transmits keep-alive beacons at the ISM band as is typical for most satellites. %The inability to geo-locate such satellites precisely is the reason why our satellites were awarded an exemption from checking geo-location first prior to transmission in the ISM band. 

\vspace{0.03in}\noindent\textbf{Antenna and Transmitter Design:} The LoRa CubeSat uses a HopeRF RFM95PW (operates similar to SemTech SX1257) transmitter with an in-built power amplifier as the RF front-end for modulating the data using LoRa modulation. This brings the output wireless transmit power to +27 dBm. The final component is to design an antenna to communicate with the ground infrastructure. An obvious constraint is the size of the CubeSat which limits our design to the 10 $\times$ 10 $\times$ 10 cm$^3$ volume. But an even more important design decision is to identify whether the antenna should be directional.

\begin{figure}[!t]
    \centering
    \includegraphics[width=\linewidth]{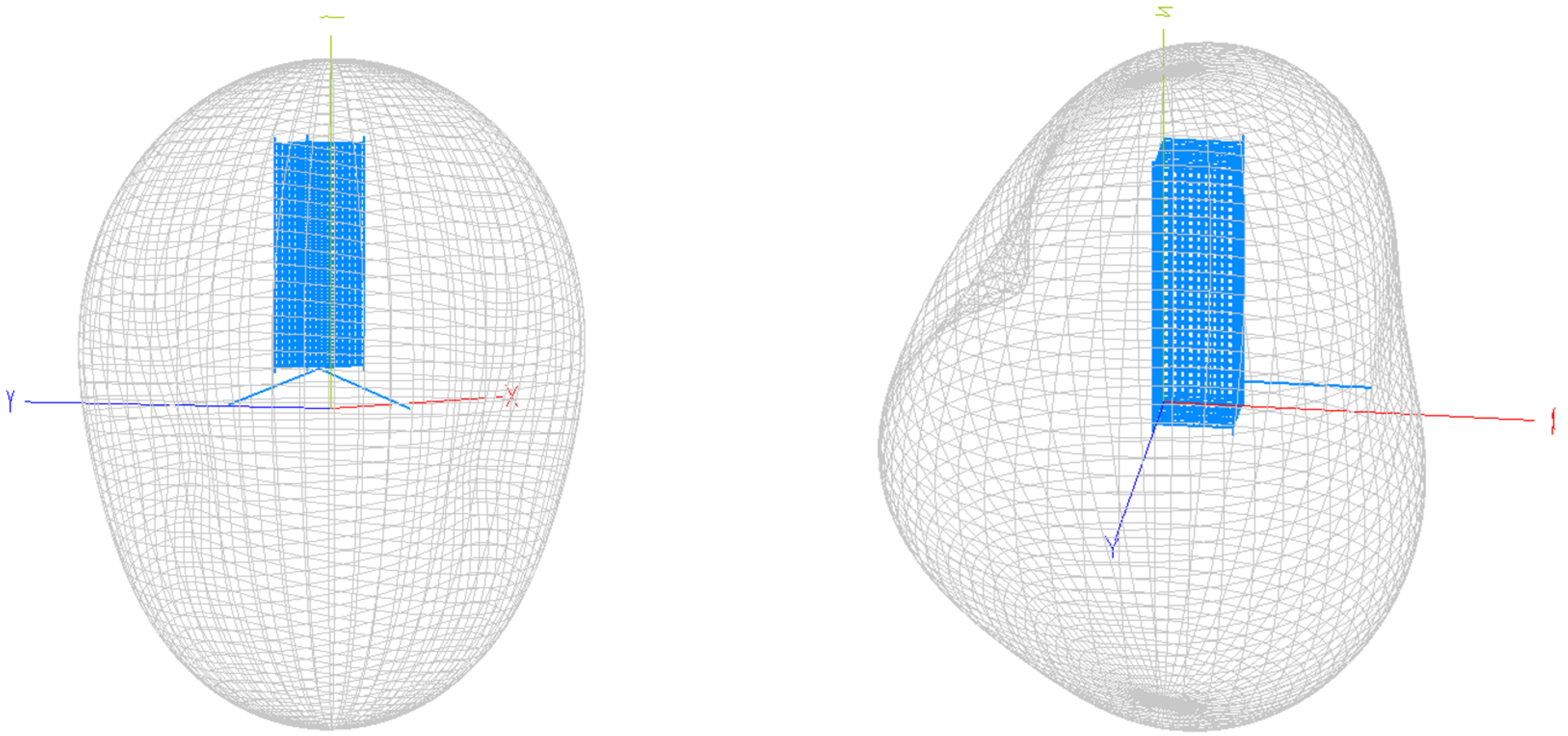}
    \caption{\name\ UHF Antenna Beam Patterns}
    \vspace{-0.2in}
    \label{fig:antenna}
    
\end{figure}

In an ideal scenario, it is clear that a directional antenna will provide additional gain improving the achievable throughput for the link. Unfortunately, small CubeSats tumble quite significantly in space -- meaning that directionality would simply be counter-productive. Further, as LoRa packets can be as long as 2 seconds, this tumbling can cause a large deviation in SNR within a single packet rendering it undecodable. Thus, the antenna on the CubeSat attempts to achieve as close to isotropic propagation beam pattern within the spatial constraints of the CubeSat while remaining tuned to 915~MHz. Fig.~\ref{fig:antenna} demonstrates the beam pattern of antennas on the CubeSat.

\noindent\textbf{Choosing Transmission Parameters:} After designing the physical hardware for the transmitter, it is critical to maximize the throughput from the link to ground infrastructure. The key factors that govern the overall throughput of the CubeSat-ground infrastructure link are -- how frequently the CubeSats transmit and amount of data in every packet.

The number of downlink transmissions is primarily governed by the energy available at the satellite throughout the day. When the CubeSat is facing the sun, it generates energy in fast-charging mode at the rate of 1W while it generates 0.5 W in trickle-charging mode when in eclipse. Given this coarse estimate of power generation and other transmission parameters, we can choose the number of transmissions per second based on the power available per packet. The satellite transmits one packet every 30 seconds in our case.

Further, the transmission power should be set to the maximum possible from the software end. The next important trade-off is to choose the spreading factor (SF) and bandwidth for the LoRa transmitter. The throughput, energy consumption and SNR coding gain of LoRa transmissions varies as:
\vspace*{-0.08in}
\[\texttt{Throughput}\propto \dfrac{BW}{2^{SF}};~~~~\texttt{Energy}\propto{BW\times2^{SF}};~~~~\texttt{SNR}\propto{\dfrac{2^{SF}}{BW}}\] \vspace*{-0.15in} 

At first glance, a simple approach of choosing the lowest bandwidth and choosing the spreading factor based on the SNR gain required looks promising. However, there is another factor that affects the packet decoding on the ground stations -- Doppler (see Sec.~\ref{sec:groundstation}). Since Doppler shift in the 915 MHz band is significantly larger than the low frequency amateur bands, low bandwidth packets are more likely go undetected. While we show later (in Sec.~\ref{sec:groundstation}), how by carefully configuring ground stations we can achieve resilience to these shifts, these detection thresholds also add an additional constraint on the transmission parameters. The CubeSats developed by our team operates in the 525 km LEO orbit operating at 915.6 MHz used SF 8 and 62.5 kHz bandwidth for communicating downlink.

\subsection{LoRa Ground Station Design}\label{sec:groundstation}
Conventional LoRa ground infrastructure (operating at the 915~MHz ISM band) is not designed to receive LoRa signals from CubeSats. Further, any approach to specialize the existing ground infrastructure would negate the latency gains of leveraging existing infrastructure. Primarily, there are two key differences between the signals received at LoRa base stations from terrestrial sensors and those from the CubeSats. First, the satellites are traveling at speeds of several km/s ($\sim$7 km/s for our satellite), much larger than any terrestrial mobile LoRa clients. Second, is the fact that CubeSats operate in the low earth orbit (300 km - 800 km altitude ; 525 km for our CubeSat) which is a significantly larger distance. At the core, the base station needs to perform two tasks:  (1) Detect and Decode the CubeSat packets, and (2) Know when the CubeSat is overhead.

\sssec{Detecting and Decoding LoRa CubeSat packets:} The attenuation of the transmitted signal at the base station in a good pass ($\sim$90$^o$ elevation peak) is roughly 150 dB at an average. This leads to an average received power of -123 dBm significantly above the detection threshold of off-the-shelf LoRa base stations (-148 dBm).  Fundamentally speaking, by leveraging the existing infrastructure (high gain antennas and LNAs), it is entirely feasible to detect and decode signals from the LoRa CubeSats. Indeed, prior deployments on hot-air balloons have demonstrated communication as far as 832 km using LoRa \cite{lorarecord}. Thus, existing LoRa infrastructure with high-gain antennas can easily overcome its range bottleneck.

\begin{figure}[!t]
    \centering
    \includegraphics[width=0.9\linewidth]{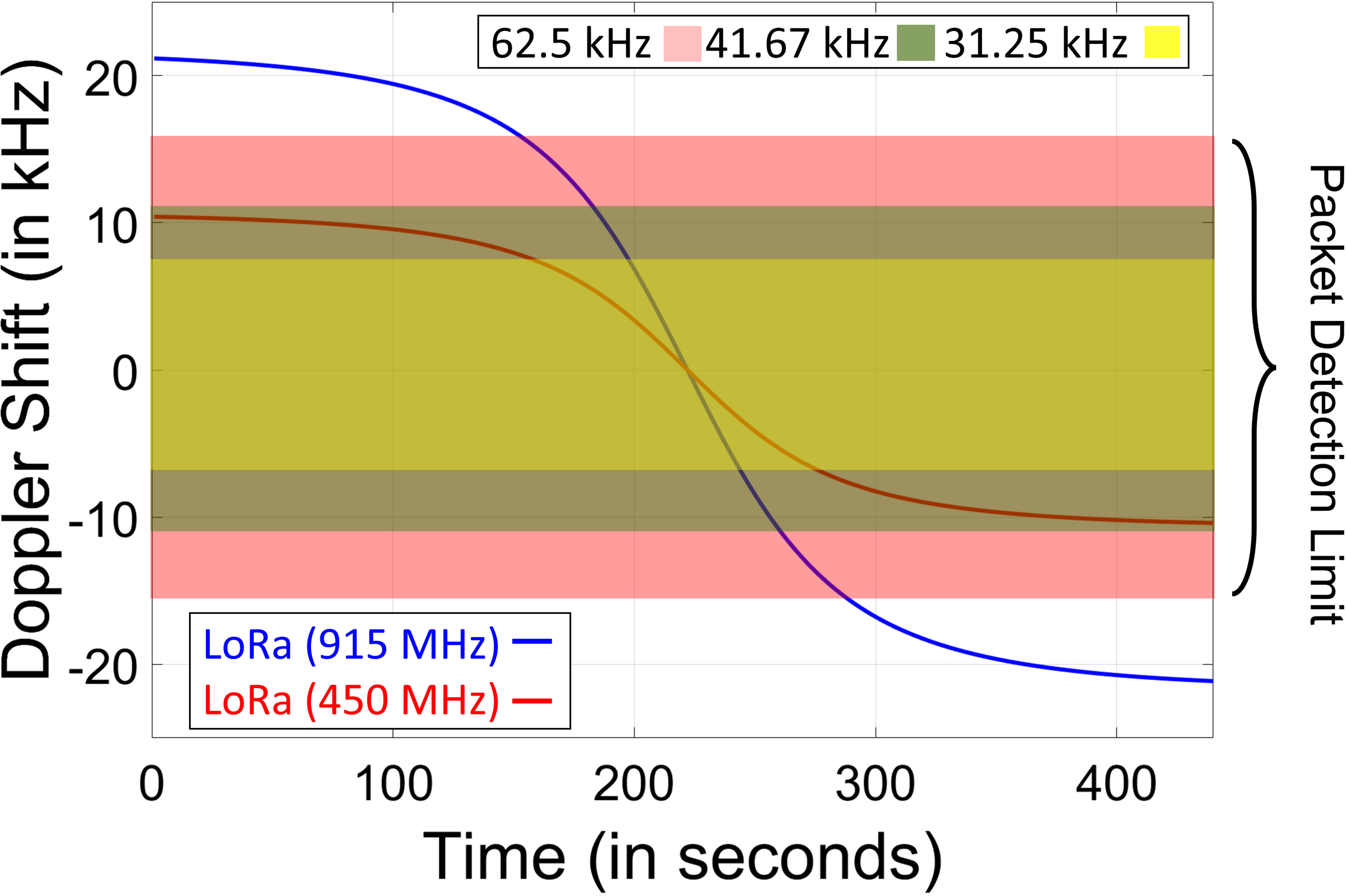}
    \vspace{-0.1in}
    \caption{Unlike CubeSats operating at lower frequencies, traditional LoRa base stations cannot cope with the amount of Doppler present in LoRa CubeSats in the ISM band}
    \vspace{-0.1in}
    \label{fig:dopplerdetection}
\end{figure}

However, the massive amount of Doppler shifts in the CubeSat context present a difficult challenge. While LoRa was developed to be inherently resilient to terrestrial Doppler~(a few Hz), the amount of Doppler seen from these CubeSats is on the order of several kHz~(a significant proportion of the bandwidth). This means up to a third of the signal might be outside the receiver bandwidth. Along with the large communication range, this renders a large fraction of the packets undetectable (see Fig.~\ref{fig:dopplerdetection}). If one were to use a software-defined radio with wide bandwidth, it would be easy to collect the raw signal data and post-process it to overcome this bottleneck. Instead, \name\ takes the approach of re-configuring existing LoRa base stations with a simple software update to estimate and overcome Doppler.

\begin{figure}[!t]

    \centering
    \includegraphics[width=\linewidth]{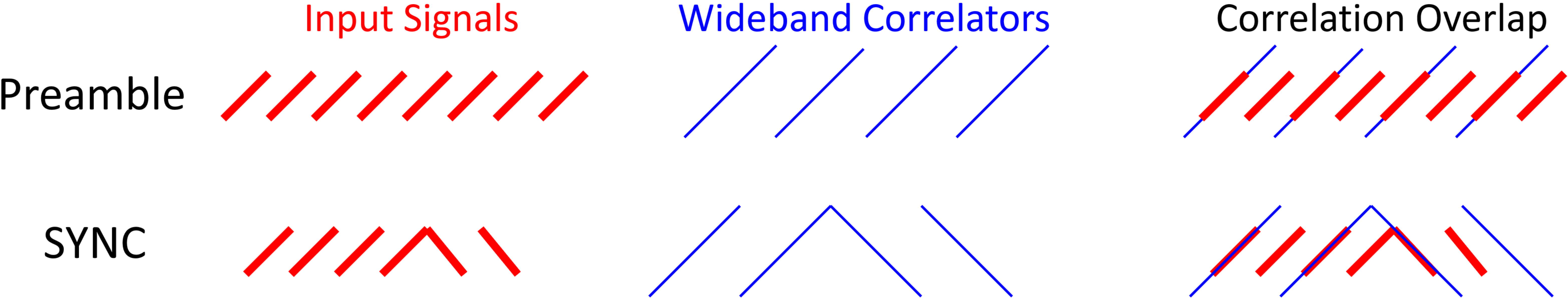}
    \caption{Correlation of preambles and SYNC symbols with wideband correlators enables high Doppler packet detection and accurate Doppler estimation }
     \vspace{-0.1in}
    \label{fig:doppic}
\end{figure}

\sssec{Packet Detection using wideband correlators:} The key reason for packet loss due to Doppler in LoRa base stations is because the correlation peak is far outside the bandwidth of the sampling unit at the receiver. To overcome this challenge, we trade-off some noise resilience for improved Doppler robustness. Specifically, we set the receiver to the widest bandwidth correlators (250 kHz) and leverage that even low-bandwidth LoRa preambles can correlate with these wideband correlators. To see why this is possible, consider Fig.~\ref{fig:doppic}: Notice that carefully configured correlators at 250 kHz share common signal fragments with those at lower bandwidths. However, only some fragments match (four out of eight preamble signals in Fig.~\ref{fig:doppic} (top)), meaning that this process results in a 3~dB reduction of detection SNR -- trading off some SNR resilience for improved detectability. At this point, \name\ can simply estimate Doppler using the SYNC symbols of a larger bandwidth (much like traditional LoRa) to obtain a coarse estimate of Doppler frequency shift. We illustrate this process in Fig.~\ref{fig:doppic} (bottom). 

\sssec{Satellite trajectory to inform Doppler and vice-versa:} As the above approach of estimating and correcting Doppler requires us to use wider preambles and accumulates more noise, it becomes critical for the base station to know when the CubeSat is overhead to switch into high Doppler-resilient mode -- trading off SNR for detectability. However, most small CubeSats cannot be tracked by ground radars for the first few weeks (out of a mission life of one year -- the most critical time for a CubeSat mission)~\cite{tleorbit} -- owing to being highly clumped up in a volume with other entities launched with them. However, the few initial Doppler measurements obtained from our LoRa ground stations can help us indirectly inform satellite orbits, indeed much earlier than ground radars can capture them. These orbital measurements can then directly inform future expected Doppler values for these same satellites in subsequent passes.

We use a simplified version of prior Doppler-to-orbit models~\cite{guier1998genesis,guier1959doppler} to estimate the trajectory from our coarse Doppler measurements using the algorithm described in Alg.~\ref{alg:trajectory}. While there are many more complex and accurate models for estimating the trajectory from Doppler estimates, our results in Sec.~\ref{subsec:to} really highlight the improvement in trajectory estimation due to better Doppler estimation by \name. Upon estimation of this trajectory, we can leverage this information in future packets to get more accurate Doppler estimates of the incoming packets. Note that we can also use this information to reduce the bandwidth of correlator symbols to improve the SNR of reception. Finally, we can use this modified Doppler estimate to correctly decode the communicated information on the ground infrastructure and leverage it for decoding images and performing tasks.

\begin{algorithm}[!t]
  \caption{Trajectory Estimation using Doppler}\label{alg:trajectory}
  \textbf{Input: Measured Doppler $f_o^1$, $f_o^2$ ... $f_o^n$, Period $t_p$}\;
  \textbf{Output: Trajectory parameters $\theta_{max}^*$, $\phi^*$, $t_{start}^*$}\;
  $// \texttt{Create Spline interpolated signal}$\;
  $//\texttt{s(t)=Spline(\{$0,t_p,2t_p$, ...\},\{$f_o^1$, $f_o^2$ ... $f_o^n$\})}$\;
  
  \nl \ForAll{$\theta_{max}\in\{0\rightarrow\frac{\pi}{2}\},\phi\in\{0\rightarrow\pi\} $}{%
        $s^*(\phi,\theta_{max})=D(t,\phi,\theta_{max})$\;
        $//\texttt{D is the emulated Doppler Curve}$
        $\Gamma(\phi,\theta_{max})=max(s^*(t)*s(t)$)\; //Cross Correlation
        $\tau(\phi,\theta_{max})=\argmax_t(s^*(t)*s(t)$) \;
  }
  \nl $\phi^*,\theta^*_{max}=\argmax_{\phi,\theta_{max}}\Gamma(\phi,\theta_{max})$\;
  \nl $t_{start}^*=\tau(\phi^*,\theta_{max}^*)$\;
\end{algorithm}

% Given these parameters, we formulate our optimization as finding the best orbital parameters that fit our Doppler observations ($DO_i$) across packets ($i= 1, \dots, N$), specifically:
% \vspace*{-0.15in}
% \begin{gather*}
%  \arg \min_{t_{start},\phi,\theta_{max}} |\alpha| \\
%  \text{where~}\alpha=\sum_{i=0}^N\Bigl|DO_i-D\Bigl(t_{start}+it_{period},\phi,\theta_{max}\Bigl)\Bigl|
%  \end{gather*}
% %2/(omega_f)*acos(cos(acos(R_e*cos(theta_v)/r)-theta_v)/cos(acos(R_e*cos(theta_max)/r)-theta_max))

% The above objective function is not easy to optimize given the complex expression of the Doppler behavior. Specifically, while the expression is convex along the $\phi$ and $\theta_{max}$ dimensions, the randomness of $t_{start}$ stops us from using gradient descent. However, since the possibilities of $\phi$ and $\theta_{max}$ are bounded, we can directly iterate on them. 

%\vspace*{-0.2in}

%% file: text/5-imageprocessing.tex
\section{Channel-Aware CubeSat Imaging}\label{sec:chaware}
In this section, we describe our physical layer approach that efficiently encodes images into the constraints of a LoRa packet (only 256 bytes each) while maximizing the accuracy of output of tasks performed at the ground.

\begin{figure}[!t]
    \centering
    \vspace{-0.1in}
    \includegraphics[width=0.9\linewidth]{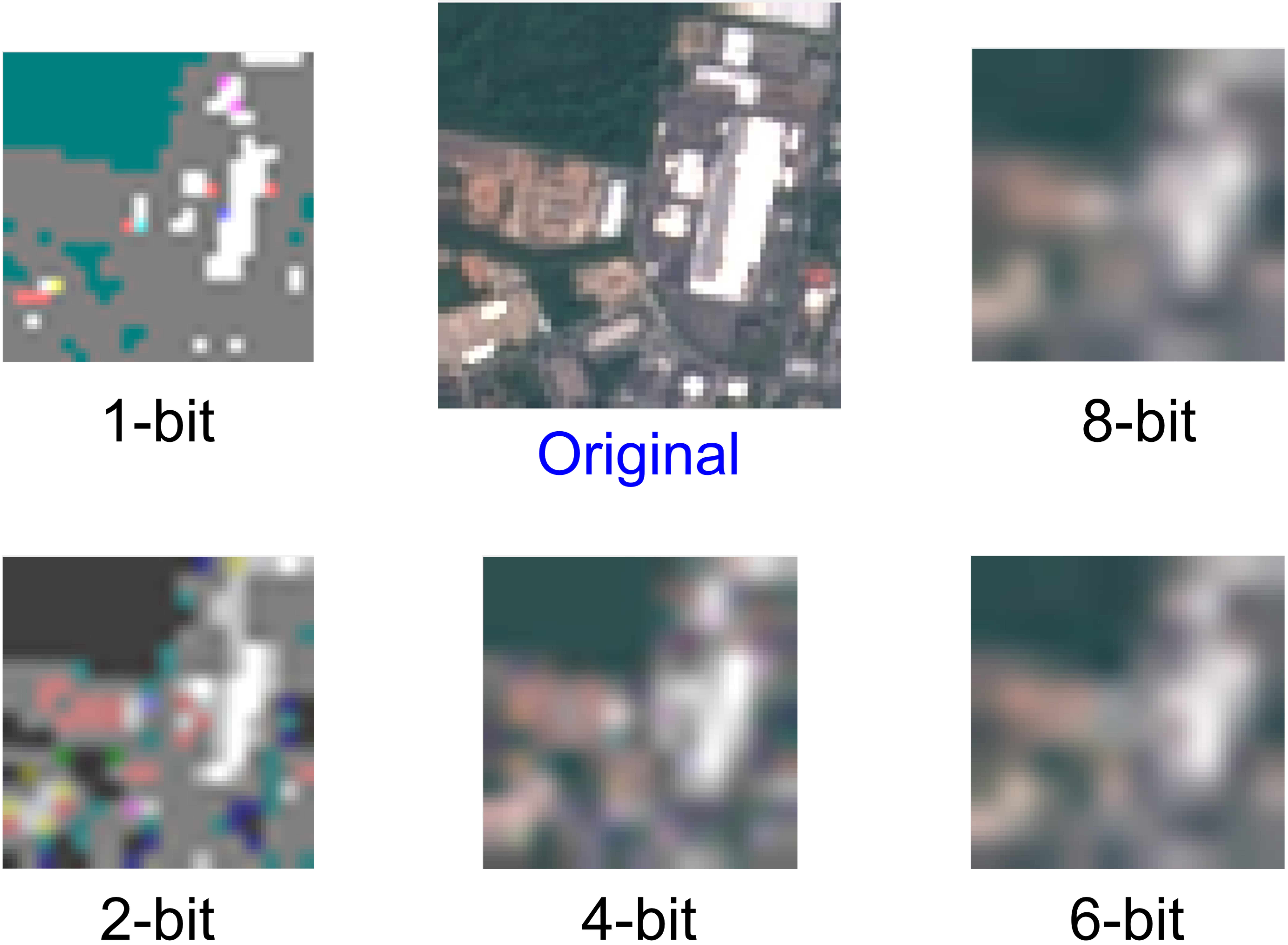}
    \vspace*{-0.05in}
    \caption{Spatial and bit-level resolution tradeoff leads to large reduction in image quality of the baseline}
    \label{fig:spatial}
    \vspace*{-0.05in}
\end{figure}

\noindent\textbf{Current approaches: }\label{sec:compressionbaseline}
Traditional approaches to compress images such as JPEG, TIFF and PNG are not designed for such low storage regimes~(hundreds of bytes). This is due to the fact that much of the compression achieved using these approaches relies on carefully designed dictionaries, hash tables and encoding schemes, all of which take space to store for decoding the image. Thus, to achieve any coherent image communication, we can only rely on two conventional compression approaches: subsampling and reducing bit resolution of every pixel. Subsampling reduces the spatial resolution of the image while the reduced bit resolution adds corruption due to discretization. Fig.~\ref{fig:spatial} demonstrates a few configurations that demonstrate significant loss of features, motivating the need for better solutions. 

\subsection{Data-Channel-Task Aware Approach}
To ensure useful data compression within the constraints of LoRa packet size, \name's key idea is to leverage important known information about the images collected by CubeSats, the wireless channel that the signals propagate over, and tasks performed on the ground. Specifically, we consider: (1) features extracted from a large dataset of typical satellite images, (2) a broad set of tasks that are typically performed on those images, and (3) a well-modelled  LoRa-CubeSat channel.

\noindent\textbf{Data Awareness:} We first rely on the vast amount of Earth images taken by larger satellites in similar orbits (NOAA~\cite{noaa}, Sentinel~\cite{sentinel}) to learn how best to compress the images. One approach would be performing a statistical analysis and identifying the patterns across images. \name\ instead chooses a more automated approach that learns these patterns using a convolutional autoencoder. 

\begin{figure}[!t]
    \centering
    \includegraphics[width=0.8\linewidth]{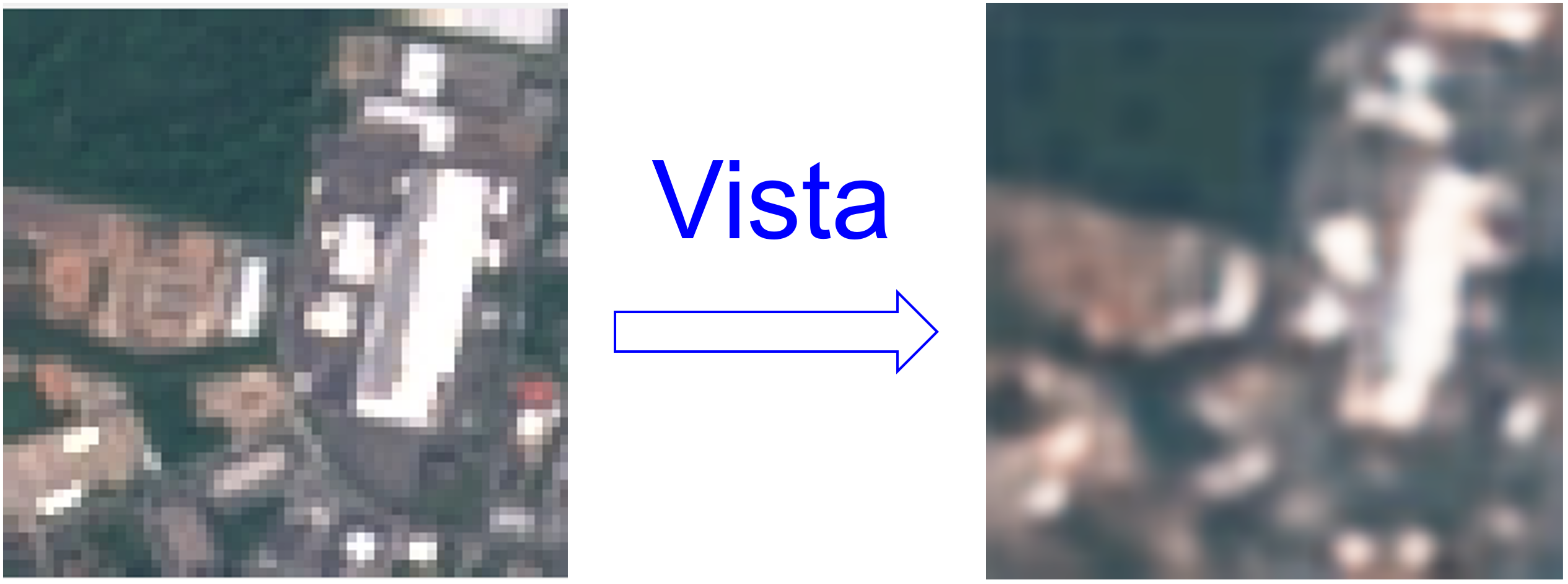}
    \caption{\name\ preserves much of the useful information in the images for inference and imaging}
    \label{fig:spatialvista}
\end{figure}

\sssec{Task Awareness:} Further, this autoencoder can be optimized not only for reconstructing the images but also directed towards preserving maximal task-specific information. We design a common encoder for both image and task-recovery that is actively aware of the set of tasks that are intended to be performed at the ground. This loses a small amount of performance in both tasks but retains maximal information feasible for each task. 

Our main rationale for designing task-aware encoders stems from a well-understood challenge in machine learning~\cite{yeh2017learning}. Specifically, certain stages of our encoder, in the process of compression, may be viewed as a series of projections onto lower-dimensional spaces. Should the data encoding be unaware of tasks performed, it may inadvertently drop critical information pertaining to these tasks. This happens due to the encoder projecting onto spaces orthogonal to those critical to our task -- which \name\ actively avoids.  

\sssec{Channel Awareness: } Finally, training and evaluating this autoencoder must account for bit-errors that these LoRa wireless packets typically encounter. We design a channel-task-aware image encoder that maximizes the output accuracy while being constrained to the limits of a LoRa packet. We do so by a two-stage process where the auto-encoder learns about \textit{noise}, \textit{intra-packet Doppler shift}, and \textit{discretization} losses. As shown in  Fig.~\ref{fig:spatialvista}, the output of our solution preserves much of the useful information in the images for inference and imaging.

\subsection{\name\ Autoencoder Design}
As we designed an autoencoder, we needed to make a decision between different types of autoencoders -- two of the common ones being convolutional (Convolutional Neural Networks) and fully connected (Artificial Neural Networks). 

\sssec{A CNN-based Design: }We choose CNN over ANN due to three reasons: (1) CNNs capture spatial features better, (2) CNN weights require less amount of data to be stored on CubeSats, or even communicated upstream (15kB vs. 24MB), (3) CNNs require significantly smaller amount of compute and storage resources during evaluation.

\sssec{Deciding CNN Depth at the CubeSat: } Next, it is critical to decide how deep the network needs to be. A general rule of thumb is that deeper networks achieve better accuracy. However, much of the computation is designed to occur on the CubeSat, which has severe computational constraints. We choose to perform three convolutional layers on the satellite to maximize the accuracy-compute tradeoff. However, on the ground, we are relatively less resource constrained, and hence can have more layers. Our architecture co-optimizes for both tasks and image reconstruction using two parallel CNNs via a fully connected layer.

Our autoencoder design (shown in Fig.~\ref{fig:systemdiagram}) builds on conventional wisdom from traditional convolutional autoencoder designs. As we go from the image towards the signature, we increase the number of convolution features captured by each layer and decrease the size of filters. This allows the \name\ autoencoder to preserve the maximum amount of information while ensuring progressive compression. After each layer, we pass the output via a Rectified Linear Unit (ReLU layer) to capture non-linear behaviors which in turn maximizes the ability of \name\ to improve over simple subsampling and other linear baseline approaches. Finally, we use a tanh layer to constrain the output between (-1 and 1). 

\begin{figure}[!t]
    \centering
    \includegraphics[width=\linewidth]{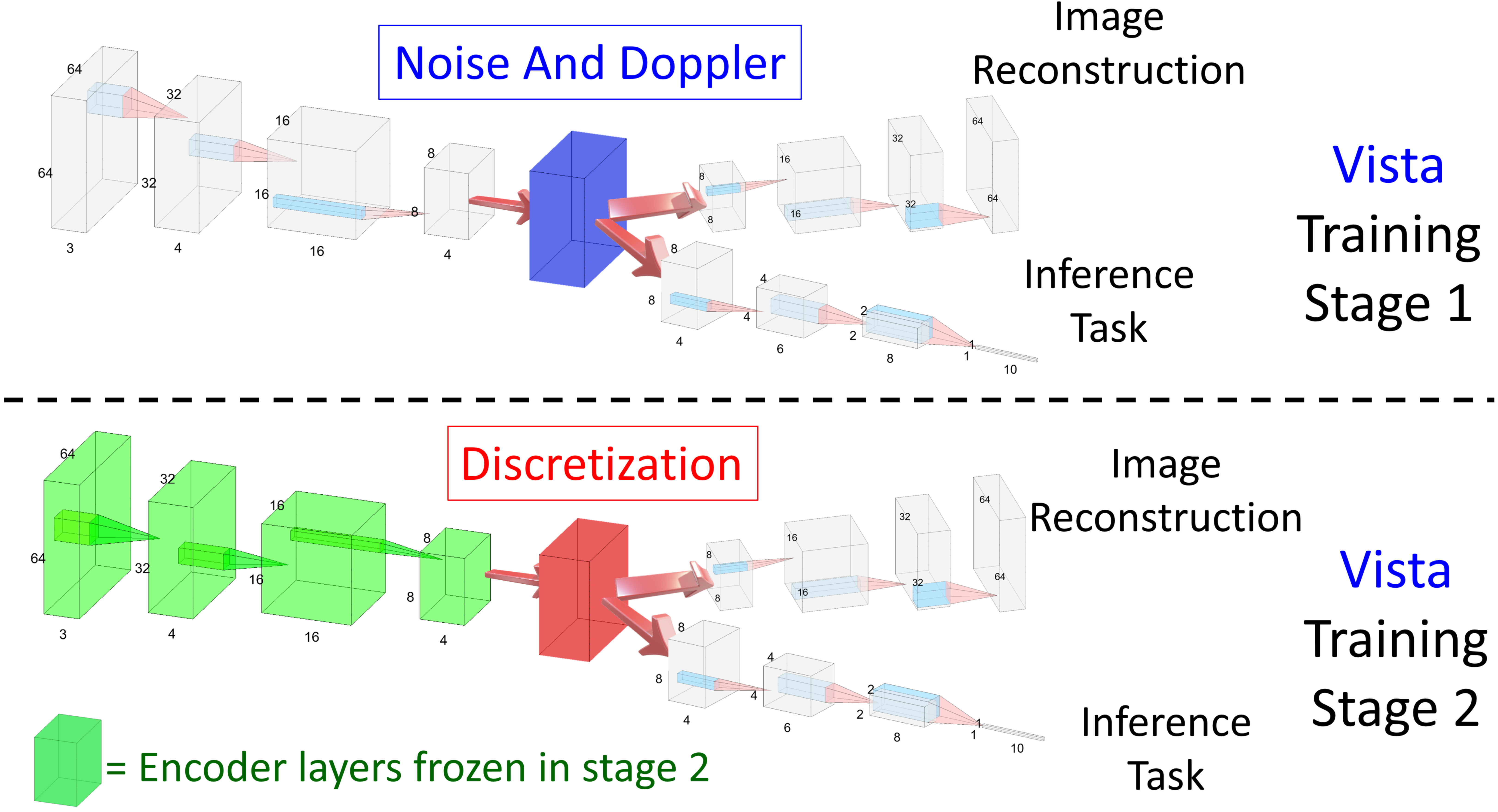}
    \caption{Vista uses two-stage training to make the autoencoder channel-aware -- \textbf{Stage 1}: learning noise and Doppler; \textbf{Stage 2:} calibration to discretization loss}
    \label{fig:channelaware}
\end{figure}

\sssec{LoRa Payload and Channel: } This output is then discretized to a low-bit representation and sent in a single packet with 256 byte payload. We note that encoder output size and discretization is primarily limited by available bandwidth. These trade-offs can be task specific, as well as dictated by the wireless channel and we discuss these issues in Sec.~\ref{ssec:channel-aware}. The packet undergoes various wireless communication impairments with two of the major factors being the noise and Doppler. Note that these CubeSats travel at 7.6 km/s at an altitude of 500 km, leading to massive Doppler offsets. At the receiver, the LoRa base station decodes the packet and retrieves the communicated values. 

\noindent\textbf{Decoder Design:} At the decoder, this signature is then passed on to a fully connected layer to distribute the input to two separate networks optimized for image reconstruction and labelling tasks, respectively. For image reconstruction, we have successive layers of upsampling, convolution, and ReLU to increase the spatial dimensions to that of the original image. Note that, during image reconstruction, we have to effectively reverse the compression process with minimal loss. To achieve this, we increase the filter size in a similar way to that of the encoder. For the labelling task, we have successive layers of convolution and ReLU layers with a final sigmoid layer to output the probability of belonging to a class.

\sssec{Training the Network: } We use the Mean Squared Error loss and Cross Entropy Loss functions for image reconstruction and multi-class classification, respectively: 

\[l_n=(x_n-y_n)^2\]
 \[\text{loss}(x,\text{class})=\text{w}[\text{class}]\biggl(-x[\text{class}] + \log\Bigl(\sum_j\exp(x[j]\Bigl)\biggl)\]

We then train our network with five-fold cross validation on the training data to avoid overfitting while ensuring maximum accuracy. As shown in Fig.~\ref{fig:channelaware}, our training occurs in two stages: In the \textit{\textbf{first stage}}, the network learns about the noise and Doppler related bit-errors. In the \textit{\textbf{second stage}}, it compensates for the discretization loss. Note that both these stages of training will occur on the ground a priori and, hence, can be performed over a long time to converge. However, despite the simple architecture, there are several design decisions taken to improve resulting accuracy.

\subsection{Channel-Aware Autoencoder Design}\label{ssec:channel-aware}

Next, we discuss how to make our autoencoder resilient to impairments of the CubeSat-LoRa Ground Station wireless channel: \textit{noise}, \textit{intra-packet Doppler shift}, and \textit{discretization}. %While wireless noise results in a uniform likelihood of bit errors across the bits, the intra-packet Doppler shift results in a non-uniform distribution of bit-errors due to the specialized LoRa chirp-spread-spectrum modulation.

%To address the problem of bit-flips introduced due to \textbf{noise}, it is critical to understand the reason behind these errors.

\sssec{Modeling noise: } Noise causes LoRa chirps to be misclassified randomly across the frequency shifts. This means that each bit has an equal likelihood of bit-flip regardless of the location of the chirp. To estimate this error, we use the LEO satellite channel model as shown in \cite{osborne1999propagation}. We estimate the symbol error rate (SER) using the theoretical model in \cite{afisiadis2019error}. Based on measurements from our LoRa CubeSats launched in 2021~\cite{vr3x}, our average SNR of -10 dB from these traces at LoRa ground stations correspond to roughly 10$^{-4}$ SER. This directly corresponds to a BER of 10$^{-4}$.

% as a function of elevation angle($\alpha$) as:

% \begin{gather*}
%     L(\alpha)=  32.44+20\log(d(\alpha)\text{km})+20\log(f\text{MHz})+L_{air}(\alpha)+\\
%     L_{rain1}(\alpha)+L_{foliage}(\alpha)+L_{fog}(\alpha)+L_{iono}(\alpha)+L_{frad}(\alpha) \\
%     \text{where } d(\alpha)=\sqrt{R_e^2+r^2-2R_er\cos{(\cos^{-1}(\frac{R_e}{r}\cos{\alpha})-\alpha)}} \\
%     R_e=6378.14\text{km},h=500\text{km}, L_{air}(\alpha)=0.1(1+\cos(\alpha))\text{dB}\\
%     L_{fog}=0\text{dB},L_{iono}+L_{frad}=2.2\text{dB}, f=915\text{MHz} \\
%     L_{foliage}(\alpha)=\bigl(7.5e^{-2.8\alpha}+0.35\cos(\alpha)\bigr)\sqrt{\frac{f\text{MHz}}{900}}\text{dB}
% \end{gather*}

% Our measurements from actual LoRa CubeSats launched in 2021 corroborate the estimated 150 dB of channel loss. We then estimate the symbol error rate(SER) using the theoretical model\cite{afisiadis2019error} build on AWGN noise as: 

% \[P(\text{\^{s}}\neq\text{s})\approx Q\Biggl(\frac{\sqrt{SNR}-(H_{N-1}^2-\frac{\pi^2}{12})^{0.25}}{\sqrt{H_{N-1}-\sqrt{H_{N-1}^2-\frac{\pi^2}{12}}+0.5}}\Biggr)\]
% By estimating the noise floor, and using trace driven measurements, our SNR of -10 dB corresponds to roughly 10$^{-4}$ SER. This directly corresponds to a BER of 10$^{-4}$ as previously stated.

\begin{figure}[!t]
    \centering
    \includegraphics[width=\linewidth]{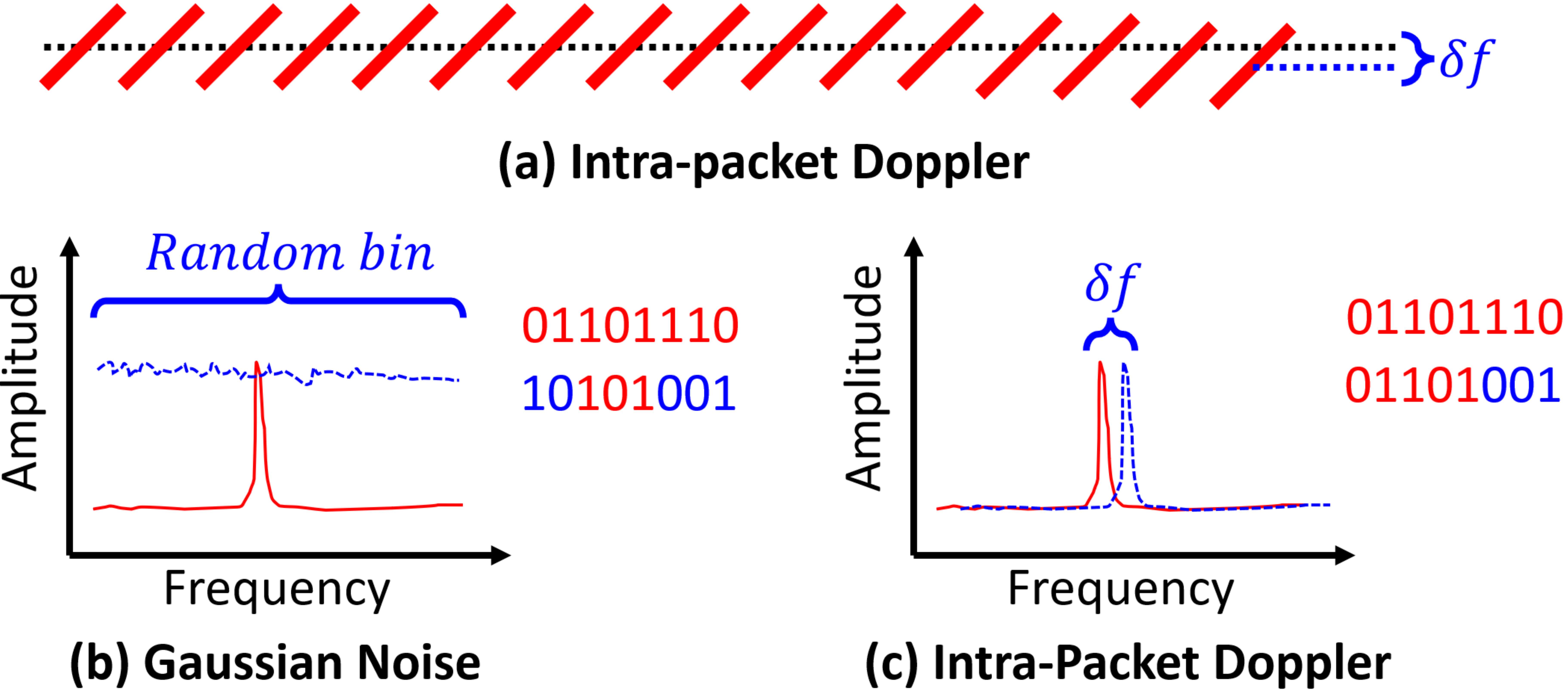}
    \caption{LoRa symbols face an additional frequency shift due to accumulation of Doppler across the packet. Unlike gaussian noise which affects all bits of a symbol uniformly, intra-packet Doppler affects the LSB more than the MSB} %(Data for 62.5 KHz SF 10)}
    \vspace*{-0.1in}
    \label{fig:doppleraccumulation}
\end{figure}

\sssec{Intra-packet Doppler: } Intra-packet Doppler affects bit errors asymmetrically. To understand how, we first need to understand how a LoRa chirp is decoded: The receiver typically deconvolves an incoming chirp with another upchirp to ascertain the frequency shift relative to the frequency shift of the preamble. Thus, generally much of the Doppler is overcome by the preamble and one would assume there would be no effect on the decoding process. While this is true for traditional wide-band satellite transmissions lasting a few millisecond, many LoRa packets last several hundreds of milliseconds. This means a significantly larger amount of Doppler accumulates over the duration of the packet, causing the relative offsets of chirps near the end of a packet to be severely deviated from the start of the packet ~(Fig.\ref{fig:doppleraccumulation}a). 

\begin{figure}[!t]
    \centering
    \includegraphics[width=0.9\linewidth]{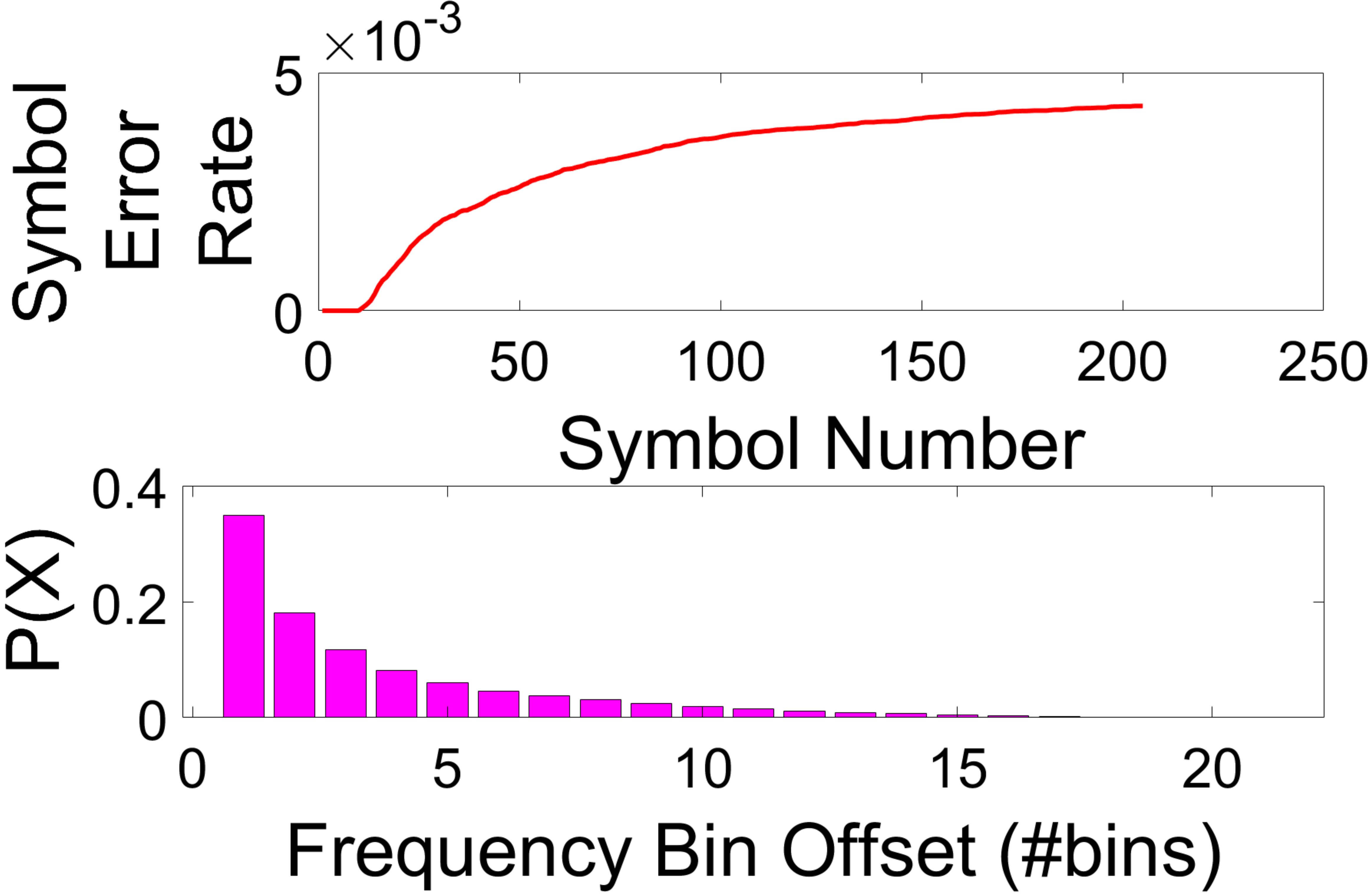}
    \caption{Intra-packet Doppler affects symbols in a packet asymmetrically. The error likelihood in frequency bins offset follows an exponential distribution.} %(Data for 62.5 KHz SF 10)}
    \label{fig:dopplerber}
\end{figure}

Doppler-induced frequency shifts lead to the decoded chirps shifting into nearby bins. This phenomenon of mis-classification into nearby bins instead of random frequency bins leads to the asymmetric effect across bits~(Fig.\ref{fig:doppleraccumulation}c). Remember that most adjacent bins only differ in the least significant bits, while the most significant bits(MSB) only change across massive frequency shifts(LSB). In fact, going from the MSB to the LSB, the likelihood of each bit getting affected increases exponentially. Thus, we need some way to inform our network model of this asymmetry. Further, simply using the approach of specializing the decoder would not work in this situation as this is critical for the encoder to understand during encoding. Fig.~\ref{fig:dopplerber} shows the distribution of symbol errors across symbols and the likelihood of amount of frequency bins shifted.

To model this effect, during the first stage of training, we add an additional layer that multiplies the input by a bit-flip operation layer. For a one bit output, we achieve this by creating random bit vectors of bit flips and then convert them to a (-1,1) vector (-1 signifying bit flip). Note that that in a one-bit output scenario, the output of the tanh layer lies between (-1,1) which means a simple multiplication layer with our bit mask will work. Further, as this bit-mask remains constant during an epoch, the multiplication layer is both continuous and differentiable. Thus, during emulation of the bit-error, the gradient will also propagate over our layer and hence get corrected automatically. Note that this makes the network aware of the behavior of these errors and maximizes accuracy for the resulting image and task. 

A new problem arises when each output is represented by multiple bits: The correct way to emulate bit flips would be to take our bit mask and XOR with the output of the encoder. However, it is well known that the XOR operation is not continuous or differentiable. To address this, we use a known approximation of the XOR operation~\cite{xorNN} by a 2-layer neural network for the number of bits required to XOR. Hence, we add this XOR black box in place of the multiplication layer where the inputs to the XOR are given by our bitmask. This enables multi-bit outputs to learn the behavior of the bit error distributions caused by intra-packet Doppler.

\sssec{Modeling Discretization Loss:} Another source of error is {discretization}. Remember that we are limited to communicating only 2048 discrete bits while the output of most neural networks are floating point integers. Thus, we need to reduce these 4 byte float output elements to a few bits causing information loss. This means that if we use a 2$\times$32$\times$32 sized output, we can only send one bit of each floating point number to the actual decoder. On the other hand, using a 1$\times$16$\times$16 output for the encoder would allow us to communicate 8 bits of information to the output. As one can imagine, both of these scenarios severely affect the ability of the decoder to decode this input into coherent output. 

To address the problem of discretization, we first attempt to understand how it affects the output of the decoder. Our observation on a well-trained CNN suggests that the discretization loss at the input of the decoder propagates to roughly equal information loss in the output. For example, the output of the 2$\times$32$\times$32 encoder becomes a single bit-mask on the output of the decoder. On the other hand, 1$\times$16$\times$16 image output suffers significantly less loss. However, this image output looks half the resolution of the image output of 2$\times$32$\times$32 image, as each of the distorted pixels in the image is much larger, spanning double the number of underlying pixels.

 \begin{figure}[!t]
    \centering
    \includegraphics[width=\linewidth]{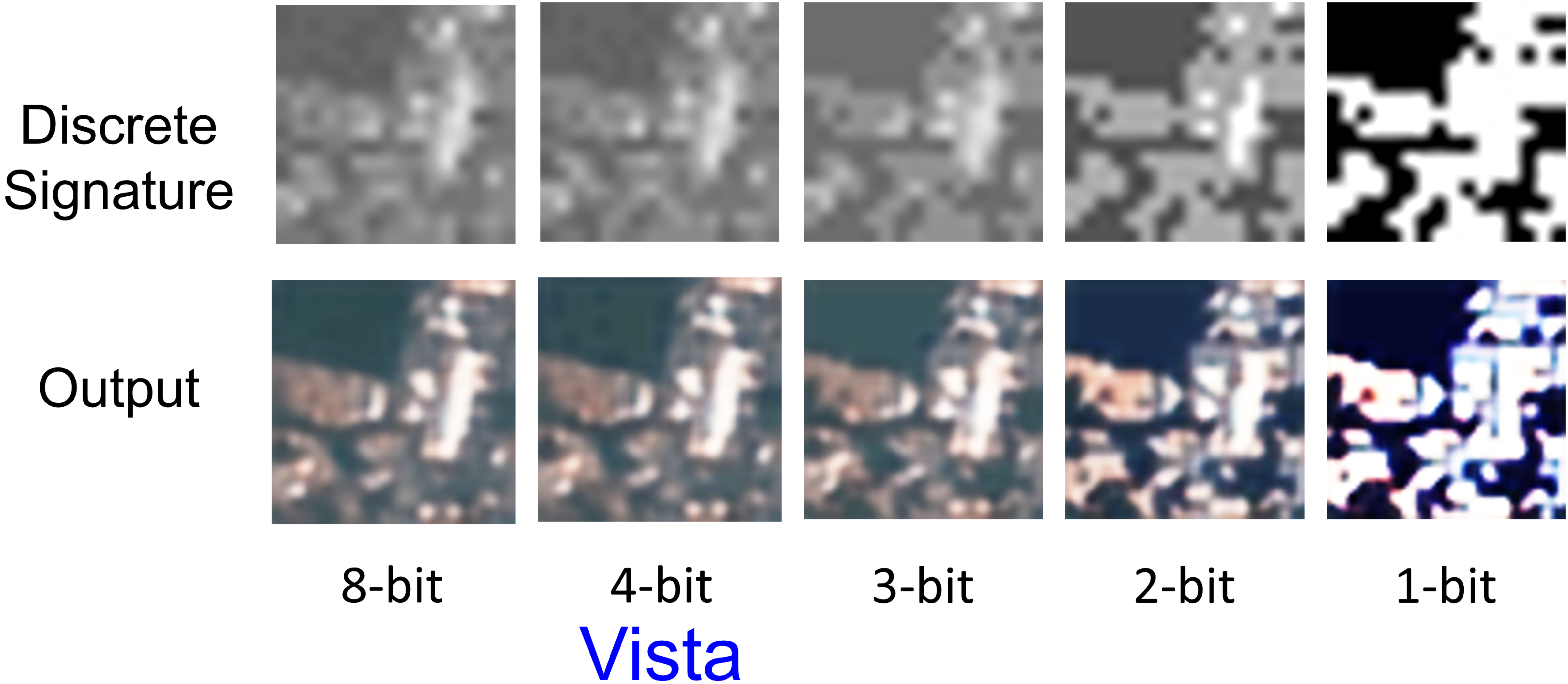}
    \caption{Discretization of the encoding propagates to the output of the decoder}
    \label{fig:discretepropagation}
\end{figure}

The average information loss for k-bit discretization is equal to $\frac{1}{2^{k+1}}$. However, it is quite difficult to quantify the amount of information lost due to reduction in spatial resolution. For instance, a lake can be just a blue blob and there will be little information lost. On the other hand, an image of an industrial complex will lose drastic amounts of information about small details. The only potential way to solve this dilemma is to evaluate the loss on the prior available data across both axes and use the least error configuration. Our evaluation shown in Fig.~\ref{fig:discretepropagation} shows that the optimum configuration n\"aively operates at 4-bits. 

However, we can do even better. We ask a simple question: ``Can the decoder learn and specialize for this discretization to maximize the accuracy of the output of the image and tasks?''  The first obviously direct approach would be to create a differentiable model of such discretization and add it as an additional layer to the neural network. However, such discretization is neither continuous nor differentiable. Further, unlike XOR, there is no guarantee that it can be estimated with a small enough neural network. Instead, we take an indirect approach to teach this behavior to our decoder: We first freeze the encoder as it is already optimized to best represent the image in the limited set of bins. We affect the output of the encoder with the necessary discretization to create an \textit{intermediate input}. We then relearn the decoder parameters for the image output on these affected intermediate inputs. Our results show how our approach minimizes information loss and maximizes the accuracy across configurations. 

\begin{figure*}[!t]
    \centering
    \includegraphics[width=0.9\linewidth]{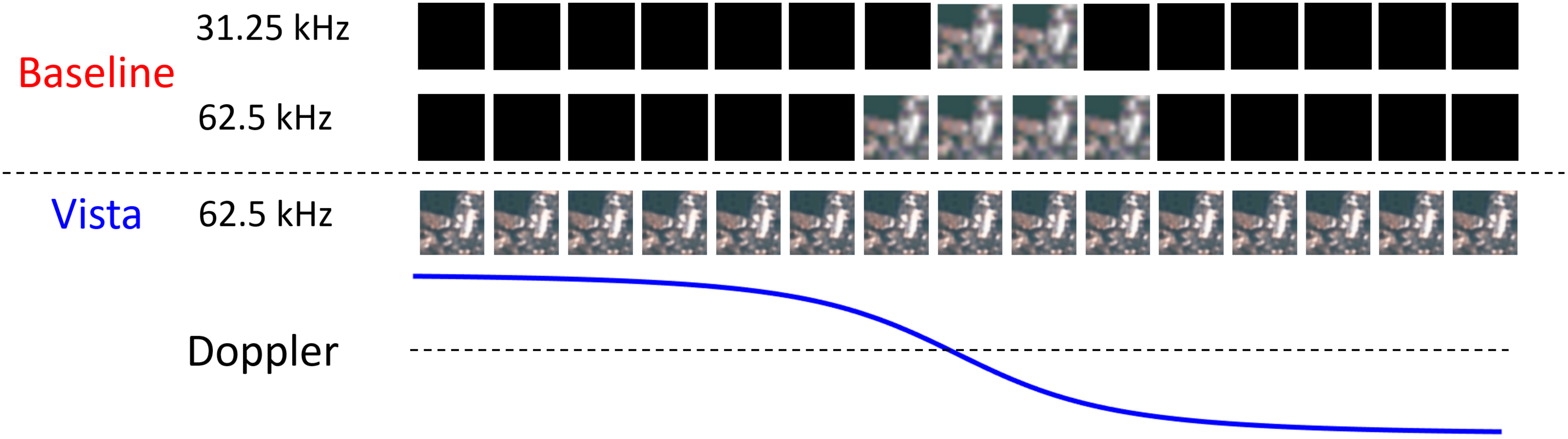}
    \vspace{-0.15in}
    \caption{Estimated image throughput over one pass of  \name\ CubeSat over an urban city with max elevation of $90^\text{o}$ and inclination angle of $97.52^\text{o}$. The baseline can communicate only 2-4 packets compared to 15 encodings of \name.}
    \vspace{-0.05in}
    \label{fig:baseline}
\end{figure*}

%% file: text/7-limitationsanddiscussions.tex
%  \begin{figure}[!t]
%     \centering
%     \includegraphics[width=0.9\linewidth]{figures/DopplerVarianceNew.pdf}
%     \caption{\name\ Satellite Doppler Variance (SatNogsDB\cite{satnogsdb})}
%     \label{fig:dopplervariance}
%     \vspace*{-0.05in}
% \end{figure}

\section{Discussion and Limitations}
\sssec{Open Challenges:} Based on our experience launching the CubeSats, it is quite evident that there are several problems that need to solved to enable robust connectivity from CubeSats to ground infrastructure. On the CubeSat,  the list of open challenges range from enabling secure robust geo-location to designing new antenna designs and modulation schemes. On the ground station, it is critical to develop new hybrid base stations that can simultaneously serve terrestrial as well as CubeSat clients and accurately estimating the trajectory of these clients in the critical first month of their operation. Finally, there is a new opportunity for data scientists to maximize the utility of these bandwidth-starved CubeSat links by building better and more robust data-driven encodings.

\begin{figure}[!t]
    \centering
    
    \includegraphics[width=\linewidth]{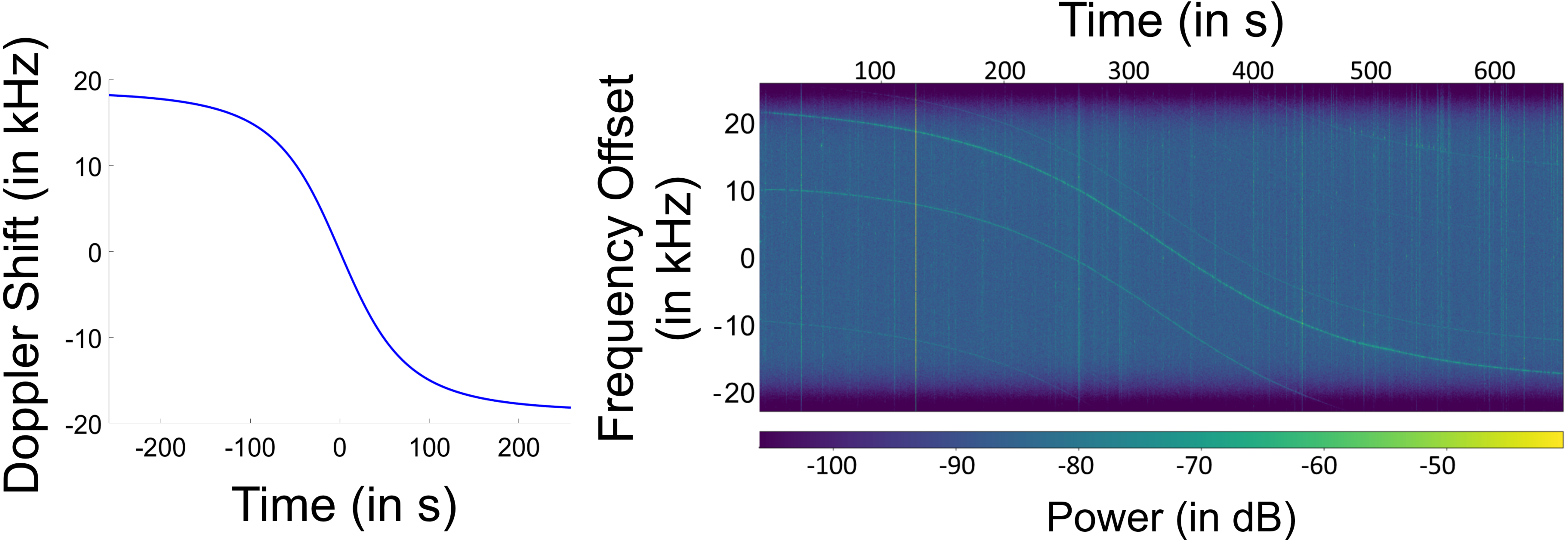}
    \vspace*{-0.1in}
    \caption{\name's evaluation leverages a model based on observed channel measurements that can replicate \name\ CubeSat channel behavior}
    \label{fig:modeldriven}
\end{figure}
 \begin{figure}[!t]
    \centering
    \includegraphics[width=\linewidth]{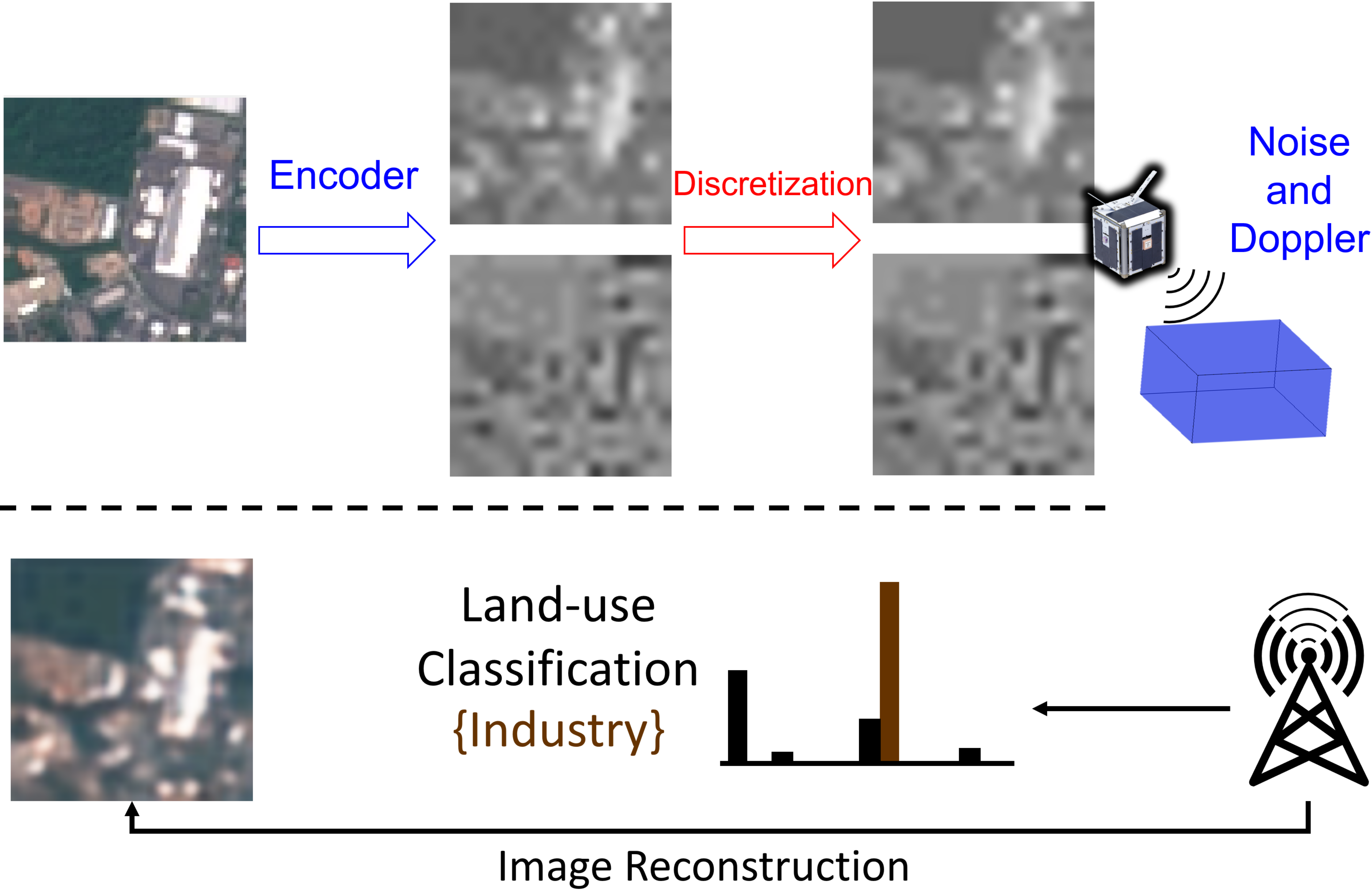}
    \vspace*{-0.15in}
    \caption{Process of \name\ evaluation}
    \label{fig:evaluation}
    \vspace*{-0.05in}
\end{figure}

\sssec{Limitations: } Our evaluation is limited in the following aspects: (1) Actual end-to-end testing on the satellite is infeasible as ground-truth image information is unavailable due to limited satellite bandwidth.  (2) Emulating high level Doppler offsets on the ground is infeasible as the orbital velocity of \name\ satellite is 7~km/s. Thus, we use the signal measurements from our launched satellite to perform trace driven evaluation. (3) We use images from public LEO satellites as candidate packet payloads for our trace-driven emulation. Note that these are not from CubeSats due to limited public data of satellite imagery specifically from CubeSats. 
%\begin{enumerate}
%    \item 
%\end{enumerate}
%
%High speed, Trace driven emulation, driven by actual problems faced on an actual satellite launch. 
%
%Need to design and test the system before actually sending up there.

%% file: text/8-implementation.tex
\section{Implementation and Evaluation}

\begin{figure*}[!t]
    \centering
    \includegraphics[width=0.9\linewidth]{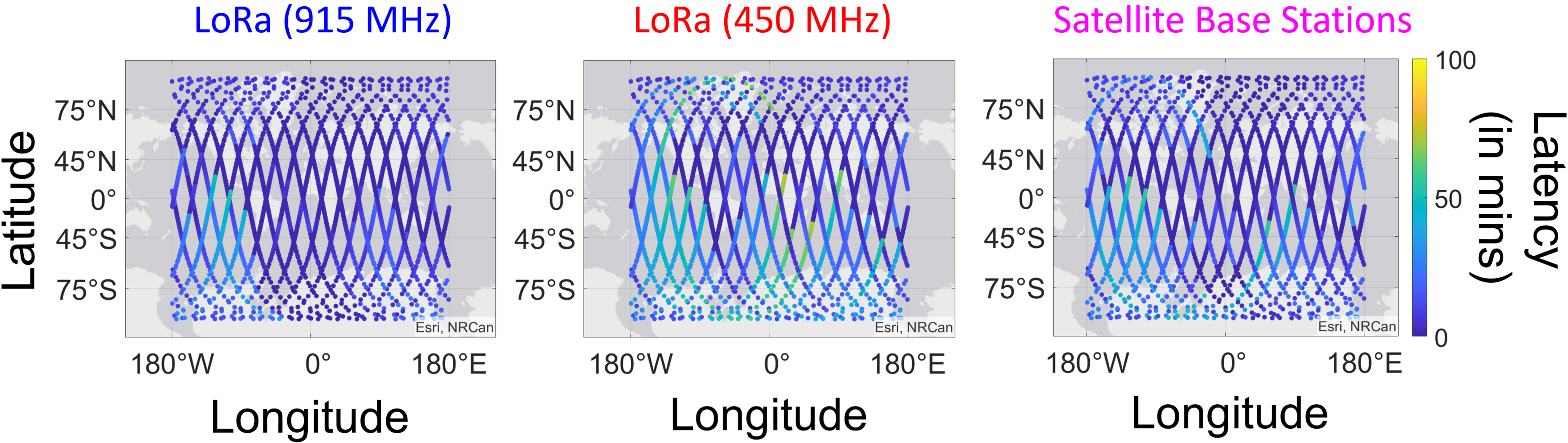}
    
    \caption{\textbf{\name\ Latency Benefits:} Demonstrates the potential latency using different ground station infrastructure over the trajectory of \name\ satellite across Feb 1, 2022 to Feb 3, 2022. LoRa ground infrastructure operating in the ISM band can enable lower latency compared to other ground station infrastructures in developing regions of the world.}
    \vspace*{-0.1in}
    \label{fig:latencymap}
\end{figure*}

Our evaluation leverages signal and channel measurements from the CubeSat to inform \name's approach to overcome the wireless challenges described in the paper. As shown in Fig.~\ref{fig:modeldriven}, our model mimics all the subtle behaviors of the satellite pass from attenuation to Doppler. We implement a ground-based evaluation testbed that emulates CubeSats attenuation and Doppler behavior using SemTech transceivers SX 1257 and SX 1262. The CubeSat orbit parameters~\cite{vr3x} were as follows: altitude = 525 km, inclination angle($\phi$) =  97.52. The anonymized TLE is below:
\vspace{-0.05in}
\[  ~~~1~--~----~-~-.-~.00096~~0~~576-2~~0~~03 \]
\vspace*{-0.2in}\[  2~-~~97.52~~88.34~~0007~~262.19~~98.73~~15.11~~16 \]
\vspace{-0.1in}

Note that the CubeSat did not receive NORAD ID up to 7 weeks after the launch due to the reasons explained in Sec.\ref{sec:loracubesatdesign}. This was a critical motivation behind our solution for identifying a way to detect the trajectory for receiving packets and useful information in the first few days of a future mission. The output power from the CubeSat was 27 dBm and the received SNR on the ground with conventional approaches varied between -10 to -30 dB. However, \name\ solutions can improve the packet detection SNR significantly by approaches mentioned in Sec.~\ref{sec:groundstation} and evaluated in Sec.~\ref{subsec:pdr}

 \begin{figure}[t]
    \centering
    \includegraphics[width=0.9\linewidth]{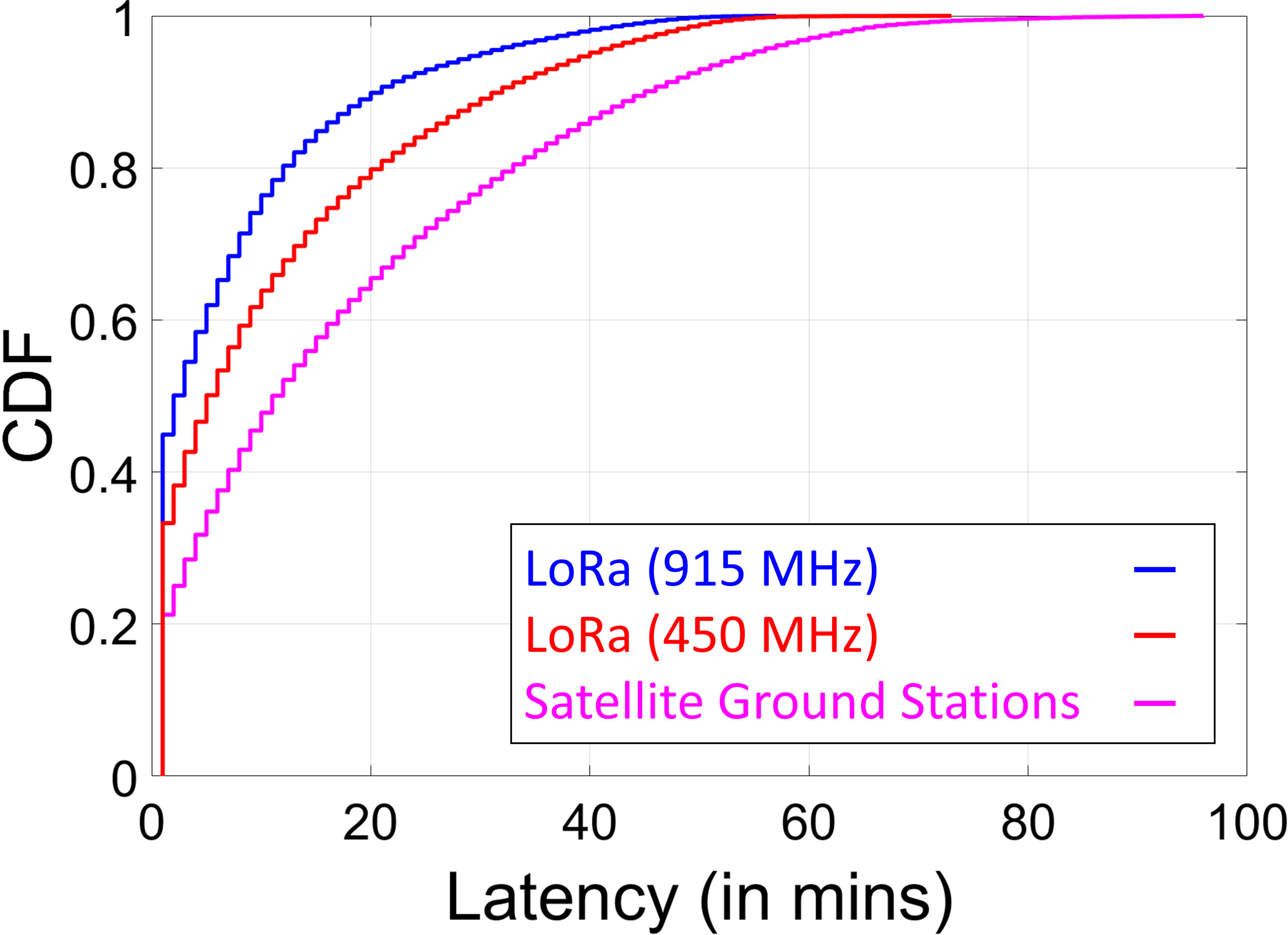}
    \caption{\textbf{\name\ Latency Benefits:} Leveraging ISM band LoRa ground stations can enable 55.55\% lower latency and 44.92\% coverage compared to 33.27\% of other infrastructure every year}
    \label{fig:latency}
\end{figure}

\noindent\textbf{Baseline:} As shown in Fig. \ref{fig:baseline}, the baseline of subsampling (Sec.~\ref{sec:compressionbaseline}) and communicating images will lose several packets without getting decoded. However, comparing our solution with pure noise would be unfair. Thus, we assume that packets were somehow detected for evaluation.

\begin{figure*}[t!]
       \centering
       \begin{subfigure}[t]{0.28\textwidth}
            \centering
            \includegraphics[width=\linewidth]{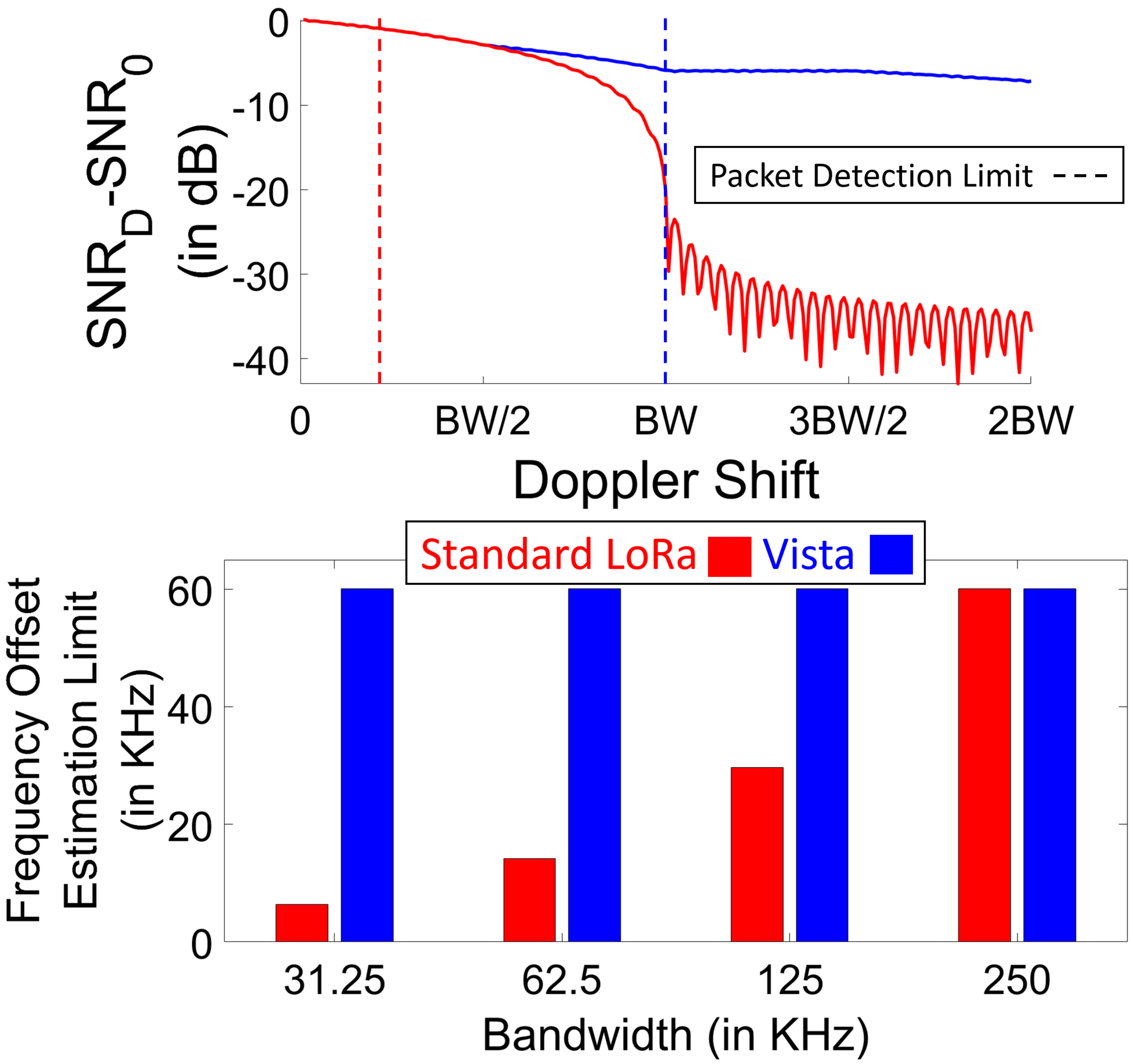}
            \caption{\label{fig:MB1}}
       \end{subfigure}%
       ~~~~ 
       \begin{subfigure}[t]{0.34\textwidth}
            \centering
            \includegraphics[width=\linewidth]{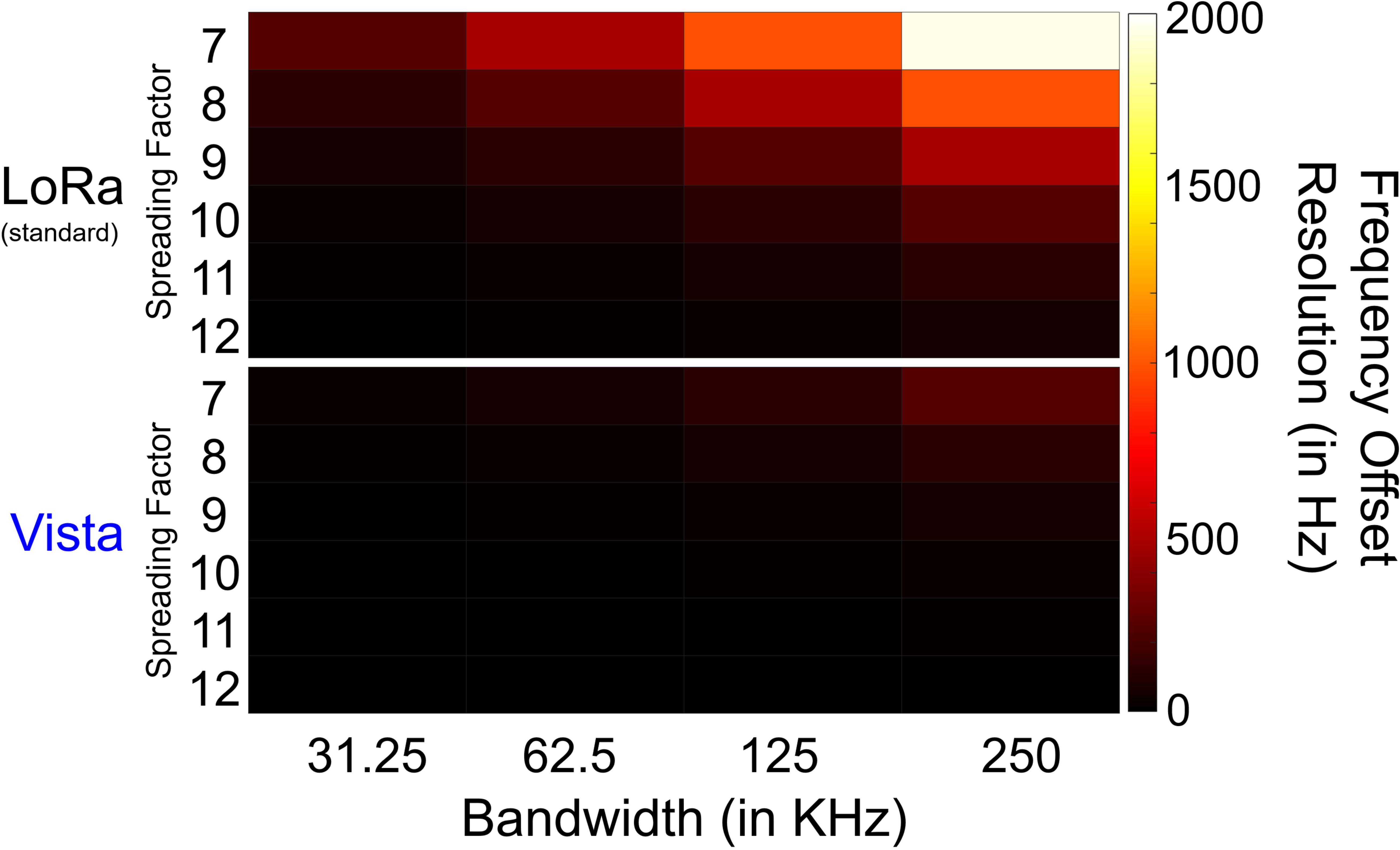}
            \caption{\label{fig:MB2}}
            
        \end{subfigure}
        ~~~~
        \begin{subfigure}[t]{0.34\textwidth}
           \centering
           \includegraphics[width=\textwidth]{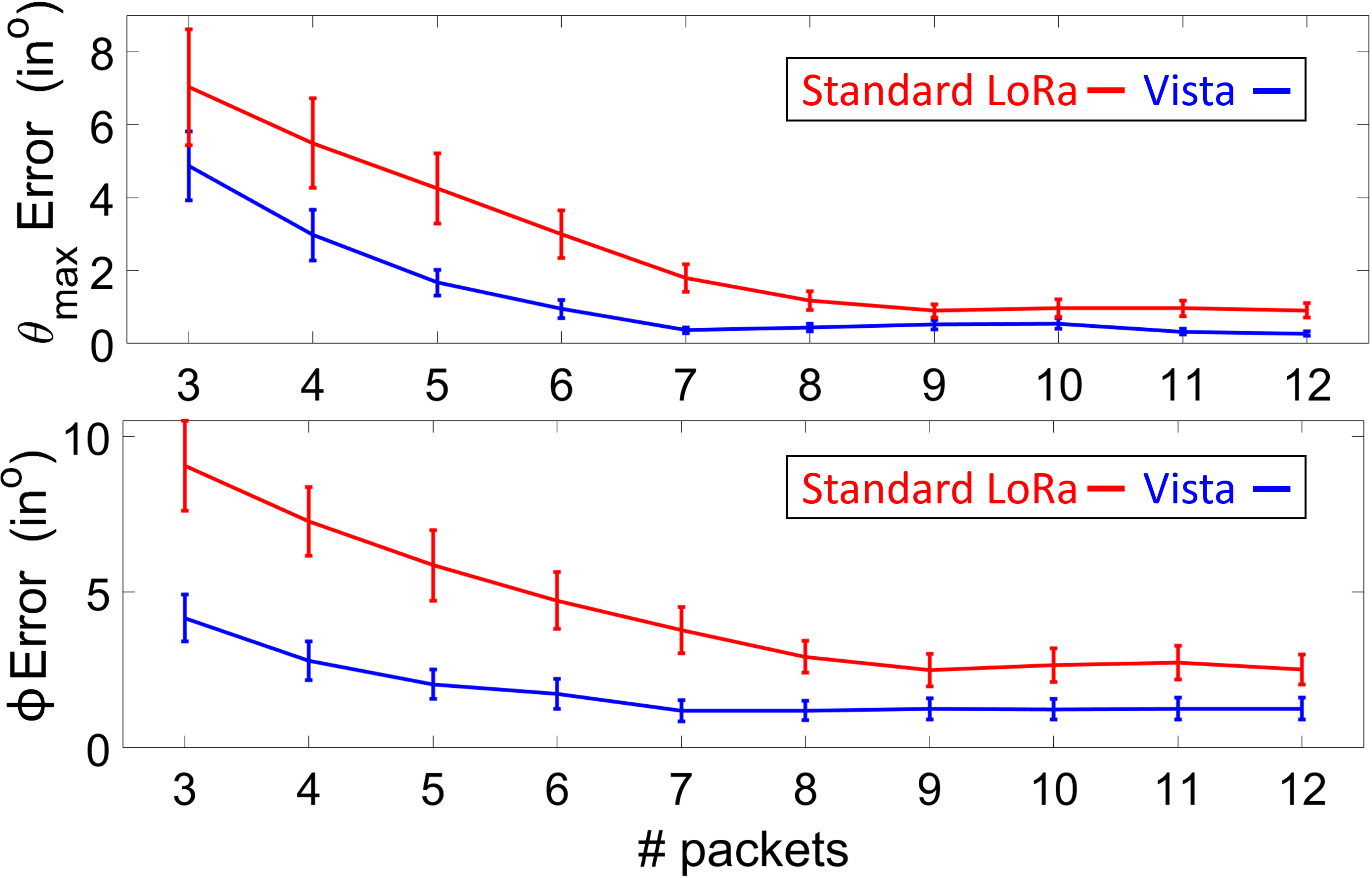}
           \vspace*{-0.16in}
           \caption{\label{fig:MB3}}
       \end{subfigure}%
       ~~~~ 
       \vspace*{-0.15in}
       \caption{\textbf{\name\ Doppler-resilient Packet Detection:} (a) Demonstrates the ability of \name\ to detect packets beyond Doppler limits of standard LoRa receivers; (b) The increased resolution of Doppler offset estimation with wideband preambles by \name; (c) \name's accurate Doppler leads to lower errors in predicting trajectory of the CubeSat}
   \end{figure*}

\sssec{Training the channel-aware autoencoder:} The training happens in two stages: (1) In the first stage during training we add the noise and Doppler layer (weights set to not change internally) and train the \name\ on training data.  (2) In the second stage we freeze the weights of the encoder and first generate \textit{intermediate inputs} as suggested in Sec.~\ref{ssec:channel-aware}. We then discretize them to fit within our communication budget of 256 bytes.  We train the decoder chain on this discretized input to calibrate for the discretization losses.

%We use 5-fold cross validation to detect overfitting and stop the training at the right epoch.

\sssec{Evaluating \name:} Fig.~\ref{fig:evaluation} shows an example run of how \name\ evaluation is performed. The input image is encoded at the encoder to output the signature which is then discretized and modulated atop LoRa. It is then passed through the channel model informed by wireless channel measurements and analyzed noise and Doppler channel in Sec.~\ref{ssec:channel-aware}. At the receiver, the signature is demodulated and passed to the decoder for outputting class labels and reconstructed image. For image reconstruction, we use the mean squared error loss and for classification, we estimate the confusion matrix. The baseline is optimally subsampled to minimize information loss as shown in Fig.~\ref{fig:spatial}. It is then passed through the same architecture without the discretization, noise and Doppler effects to ensure ceteris~paribus comparison.

\sssec{Datasets used:} We use publicly available labeled dataset with images captured from Sentinel-2, 
EuroSat Land Use Dataset~\cite{helber2019eurosat}: 27000 13-channel 64$\times$64 images with 10 seperate land use labels. For our evaluation, we restrict ourselves to RGB channels as those are the ones that image captures.

%\swarun{More details please} 

%We evaluate our system on orbit of our satellites with images captured from another larger satellite on the same orbit and perform tasks that these images are used for XXX.

%We also present a proof-of-concept evaluation based on Lunar and Martian Doppler parameters based on altitude and inclination of LRO and MRO on publicly available dataset for rockslide detection. 

%% file: text/9-results.tex
\section{Results}
In this section, we begin with evaluating two basic primitives behind \name: packet detection rate (Sec.~\ref{subsec:pdr}) and trajectory optimization (Sec.~\ref{subsec:to}) on LoRa hardware. We then present our main results for the end-to-end \name\ system based on the Sentinel-2 images for image reconstruction (Sec.~\ref{subsec:ir}) and multi-task optimization (Sec.~\ref{subsec:mcc}) driven by traces measured from our recent satellite launch.

\subsection{Latency Benefits} \label{subsec:latency}
\sssec{Setup: } We evaluate the potential latency benefits of using existing LoRa ground stations (operating on the 915 ISM band) over specialized ground infrastructure such as large generic satellite base stations or LoRa ground stations in lower UHF bands. Now, it is virtually impossible to know the locations of every possible ground station in all of these categories and thus we endeavor in procuring the publicly known locations of these ground stations on three platforms : The Things Network\cite{things-network} (terrestrial LoRa network), TinyGS\cite{tinygs} (LoRa grounds stations operating at lower bands), SatNogsDB\cite{satnogsdb} (Satellite Ground Stations). Assuming the TLE information above obtained from NORAD, we use the location of our satellite across a year in space to evaluate the potential latency of retrieving information from any of these networks.

\sssec{Results:} Fig.~\ref{fig:latencymap} shows the latency across trajectory of the satellite using various ground infrastructure during two days of operation. It is quite evident that across many developing regions (e.g. West Africa, Greenland, Indian Ocean Islands) where it is difficult to develop and deploy expensive specialized ground infrastructure, leveraging existing LoRa deployments in ISM bands provide a clear win. Our evaluation across a year of \name\ trajectory (shown in Fig.~\ref{fig:latency}) demonstrates that that leveraging the widely deployed 915MHz LoRa ground infrastructure will enable significantly lower latency of image downlink compared to using generic satellite ground stations or specialized LoRa infrastructure. Across the orbit of an year, LoRa infrastructure at 915 MHz provides a coverage of 44.92\% compared to 33.27\% of TinyGS and 21.23\% of SatNogsDB. In terms of 90\%ile latency, using \name\ for LoRa CubeSats can potentially reduce latency by up to 55.55\%.

\subsection{Packet Detection Rate} \label{subsec:pdr}
\sssec{Setup: } We evaluate the efficacy of \name's wider correlators in detecting Doppler offset packets. We use off-the-shelf Semtech SX1262 base stations to receive LoRa packets with increasing Doppler offsets and measure the limit of packet detection limits. We also measure the received signal power across offsets using SemTech SX1257 base stations (with access to I/Q samples) for \name's wider correlators over standard narrowband decoding.

\sssec{Results:} Our measurement study in Fig.~\ref{fig:MB1} demonstrates that the frequency offset estimation limit of our wideband correlators is as good as the widest correlators supported by the hardware. Further, using these correlators adaptively shows large gains in SNR reaching up to 15dB when Doppler is as large as the bandwidth. Further, packet detection limits quadruple the resilience to Doppler shifts compared to baseline LoRa base station. As shown in Fig.~\ref{fig:baseline}, this could lead to multiple-fold increase in packet throughput when receiving signals from a single CubeSat pass. Note that additional interference may reduce the gains but \name's approach will always be superior to native LoRa receivers.

\subsection{Accuracy of trajectory optimization}\label{subsec:to}

\sssec{Setup: } We use the state-of-the-art Doppler-to-orbit models~\cite{guier1998genesis,guier1959doppler} to emulate the behavior for LEO trajectory in terms of Doppler, we vary the trajectory across values of maximum elevation ($\theta_{max}$), inclination angle ($\phi$) and start time of transmission ($t_{start}$). We also add additional frequency noise as expected from a real satellite due to angular velocities, offset estimation resolution and other factors. We then use \name's approach to estimate the trajectory using Alg.~\ref{alg:trajectory}.

\sssec{Results:} Our results show that using the preamble along with the sync symbols achieves 4$\times$ better frequency offset estimation resolution compared to only two sync symbols (see Fig.~\ref{fig:MB2}). Further, due to the better resolution in measuring the offset estimate, we can estimate the trajectory with a resolution of 1.2$^\text{o}$ in inclination angle ($\phi$) and 0.5$^\text{o}$ in max elevation ($\theta_{max}$) which is comparable to the accuracy of available TLE files for much bigger satellites~\cite{tleorbit}(see Fig.~\ref{fig:MB3}). Further, the inclination angle remains consistent within a margin of 1$^\text{o}$ over the lifetime of a CubeSat, which can be further used to refine Doppler measurements.

%\subsection{Stagewise results of the autoencoder}
\subsection{Accuracy of Image Reconstruction}\label{subsec:ir}

\sssec{Setup: } Our model (Fig.~\ref{fig:systemdiagram}) uses a fully-connected layer to co-train two parallel CNNs optimized for image reconstruction and multi-class land-use classification with 50-50 (5120 images each) train-test split. The dataset is taken from~\cite{helber2019eurosat} which contains labelled images captured from Sentinel-2~\cite{sentinel} satellite in 13 different spectrums from an altitude of 800km. However, recall that a CubeSat will likely not have a high resolution camera as well as multi-spectrum sensing capabilities. We hence choose only the RGB spectrum inputs to classify the image into the 10 labelled land-use patterns: Sea \& Lake, Annual Crop, Forest, Herbaceous Vegetation, Highway, Industrial, Pasture, Permanent Crop, Residential and River. For the \textbf{\textit{baseline}}, we leverage a subsampled image with dimensions 3 channels 9$\times$9 pixels with 8 bit resolution as input. For the \textbf{\textit{data-aware approach}}, we specially train a \name\ autoencoder specifically for reconstructing images with a waist of 2 channels 16$\times$16 pixels with resolution of 4 bits per pixel. Finally, \textbf{\textit{\name's approach}} co-trains two parallel CNNs optimized for image reconstruction and multi-class land-use classification. %While we present the testing results for land-use classification later in Sec.~\ref{subsec:mcc}, this subsection only presents results from the testing data for image reconstruction.  

\begin{figure*}[!t]

       \centering
       \begin{subfigure}[t]{0.39\textwidth}
            \centering
            \includegraphics[width=\linewidth]{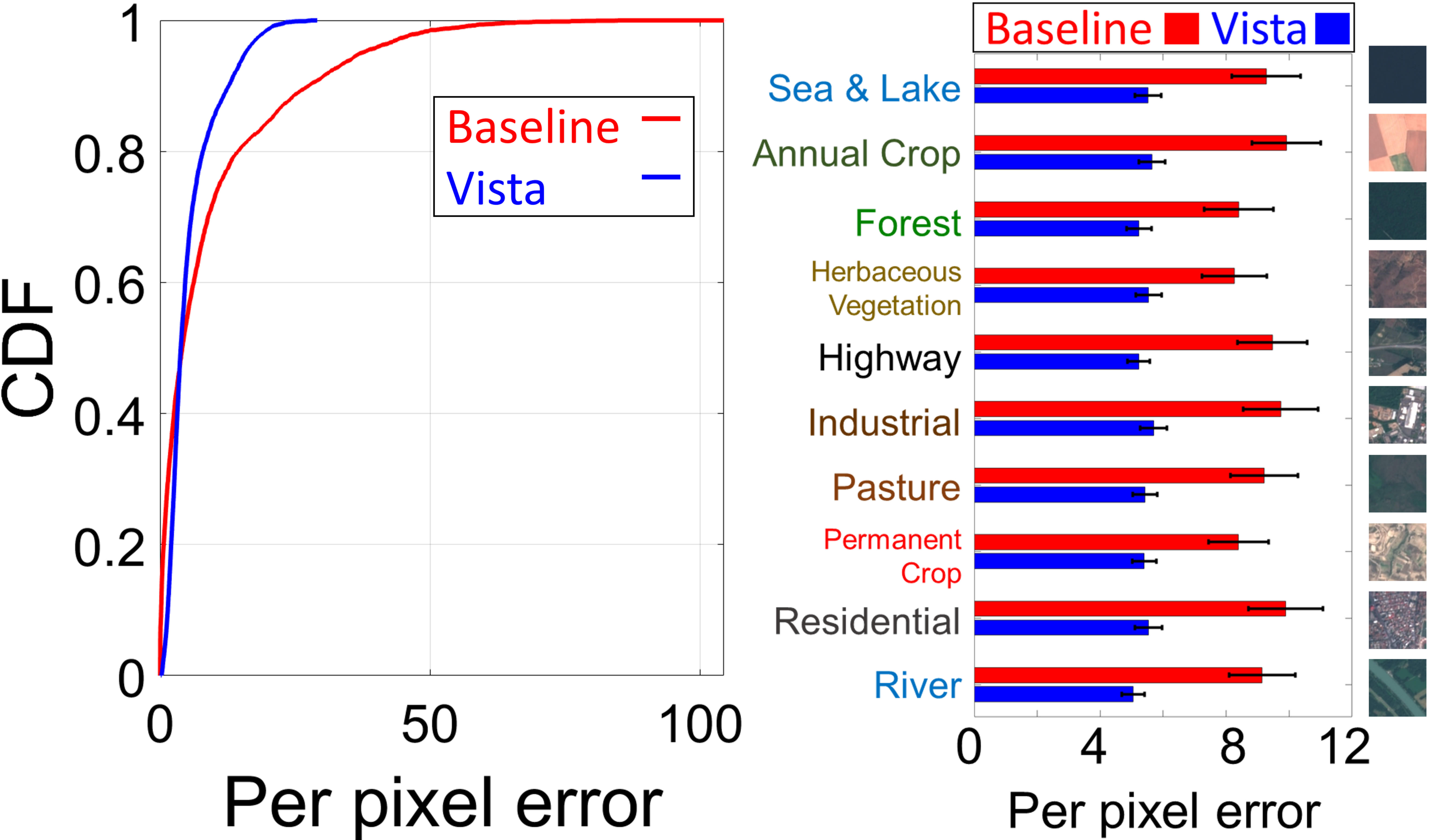}
            \caption{\label{fig:R1}}
       \end{subfigure}%
       ~
       \begin{subfigure}[t]{0.59\textwidth}
            \centering
            \includegraphics[width=\linewidth]{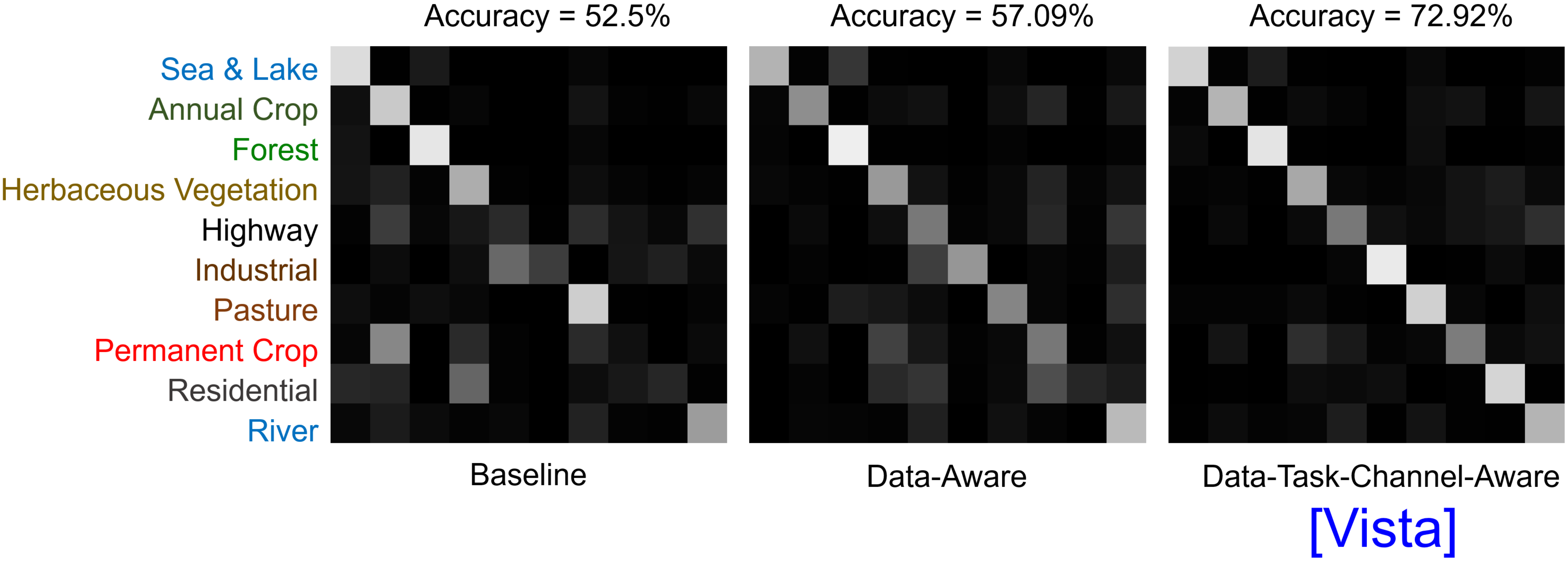}
            \caption{\label{fig:R2}}
        \end{subfigure}
        ~
     
       \caption{\textbf{\name\ Channel-Aware Imaging and Inference:} (a) \name\ produces significantly lower per-pixel error across 5120 RGB images captured by Sentinel-2 satellite leading to 4.56~dB improvement in PSNR (b) The data-task-channel aware approach of \name\ achieves 1.38$\times$ higher accuracy in classifying 10 different land-use labels compared to baseline approaches.\label{fig:Result}}
       
   \end{figure*}

\sssec{Results:} Fig.~\ref{fig:Result} shows that the baseline's per pixel error is significantly larger than Vista's approach. One of the crucial characteristics to observe is that the $90^{\text{th}}$ percentile error is especially larger for the baseline which means while in some cases the baseline may do well, as more fine grained features get involved the reconstruction becomes much worse. For example, a lake may be easier to identify from even a low-resolution image of the baseline while an industrial complex is far more challenging to make out. This lower mean squared pixel error of 5.42 (compared to 9.17)  leads to an average 4.56 dB improvement in image SNR. Note, however, that \name's approach was not specialized for image reconstruction but also for accomplishing task using the same 256 bytes of information. A specialized network such as data-aware approach which is specialized for image reconstruction may perform even better at image reconstruction but will perform poorly at tasks. This is due to the fact that the latent space of compression is not necessarily tuned in the right dimensions for the task. Thus, we do not consider this method as a fair comparison with our system for image reconstruction.

\subsection{Accuracy of Multi-Class Classification} \label{subsec:mcc}
\sssec{Setup: } Our setup remains identical to Sec.~\ref{subsec:ir} above, with the testing performed for land-use multi-class classification.  

\sssec{Results: } As shown in Fig.~\ref{fig:R2}, \name's approach outperforms both the baseline approach as well as data-aware approach significantly leading to 72.92\% accuracy in classifying the images compared to 52.5\% and 57.09\% accuracy of the former over 5120 test images (512 images for each class). The classification tasks where \name's approach achieves much of its gains are those which require high resolution intricate details such as outlines of buildings for residential and industrial complexes (higher spatial resolution). Further, industries and highways are both made of concrete (grey colored) and are hence confused much more in the baseline. The classification error that \name\ makes are primarily between structures like rivers and highways that both represent narrow lines between land or between permanent crop and vegetation. Note that a fine-tuned 13-channel 2-layer CNN (roughly comparable to our 4-layer narrow waisted CNN) achieves 84.48\% classification accuracy in the original work~\cite{helber2019eurosat} from which this dataset was obtained. But since we use only three channel RGB, our classification accuracy is lower than it.

%% file: text/2-relatedwork.tex
\section{Related Work}
In this section, we elaborate on the extensive related work done in the field of satellite communication, LoRa communication, and various image compression schemes.

\sssec{LoRa CubeSats:} Past work has explored improving the range~\cite{dongare2018charm,gadre2020frequency}, throughput~\cite{gamage2020lmac,sun2021partial,yang2020joltik}, scalability~\cite{gamage2020lmac,gadre2020quick,shahid2021concurrent, hessar2019netscatter} and reliability~\cite{afisiadis2019error} of LP-WANs. Prior systems~\cite{nardello2019camaroptera} have also delivered images over LoRa by communicating images over several packets. However, CubeSats don't have the luxury of perennial coverage. 

The idea of using LoRa for CubeSats has been explored by both, companies such as Swarm Technologies~\cite{swarm} and Lacuna Space~\cite{lacunaspace} to provide global IoT connectivity, and hobbyists as a simple off-the-shelf technology for communication~\cite{klofas2008survey} over long ranges. Much of this prior work relies on dedicated ground stations\cite{tinygs} that operate at lower frequencies (such as 137 MHz and 450 MHz) instead of traditional LoRa base stations that operate in the ISM band. Further, while there is some theoretical analysis on understanding LoRa behavior in CubeSat scenarios, most of these miss important aspects that become apparent when the CubeSat is actually deployed such as Doppler\cite{fernandez2020assessing}, ignore PHY layer considerations \cite{palattella2018enabling},  As highlighted in Sec.~\ref{sec:groundstation} and \ref{sec:chaware}, there are two key issues that particularly hinder enabling connectivity for LoRa satellites in the ISM band -- (1) Packets go undetected due to limitations of LoRa ground infrastructure (Fig.\ref{fig:dopplerdetection})~\cite{sx1262, sx1262calculator, ameloot2021characterizing,lapapan2021lora,doroshkin2019experimental} and (2) Intra-packet Doppler generates heterogeneous errors that are atypical of wireless channels. \name\ provides new solutions informed by our CubeSats channel traces in the ISM band (Sec.~\ref{sec:dop}) to overcome these limitations.

%While SemTech LoRa documentation~\cite{sx1262} and evaluation~\cite{sx1262calculator, ameloot2021characterizing,lapapan2021lora} shows resilience to small Doppler offsets, even the documented limits are lower than the actual Doppler shift present at the CubeSats. \name\ considers varied problems in deploying LoRa in the ISM band in space, including .

\sssec{CubeSat Communication and Satellite Imaging:} There has been much work on development of satellites for various applications spanning agriculture, atmospheric monitoring and remote sensing. On one end, are publicly funded programs such as NOAA~\cite{noaa}, Sentinel~\cite{sentinel}, LandSat~\cite{landsat} with large historical data, where academics have recently shown new opportunities\cite{singh2022selfiestick,singh2021community} in developing novel ground architectures for receiving image data. On the other hand, are private enterprises entering this space such as Planet Labs~\cite{planetlabs}, Microsoft~\cite{vasisht2021l2d2,vasisht2020distributed} and Boeing Aerospace~\cite{boeing}. There has been also much work in characterizing the trajectory~\cite{bohra2009analysing,shen20193} and channel~\cite{osborne1999propagation,khalife2020blind} of these LEO satellites. All of these above deployments are big satellites or large CubeSats ($\geq$ 6 unit), and leverage high throughput wideband technologies to communicate to ground infrastructure.   \name\ addresses the unfortunate consequences of using narrowband technologies in LEO -- the lower data rate as well as high Doppler-to-bandwidth ratio. Further, small CubeSats \cite{vr3x} face the unfortunate issue of tumbling in space which does not affect the larger satellites due to compensation mechanisms~(Sec.~\ref{sec:dop}).

\noindent\textbf{Error-Resilient Image Compression over wireless links:} Imaging is one of the most important applications for today's CubeSats with usecases in weather monitoring, satellite imaging, cloud detection, agriculture, etc. ~\cite{agrawal2019machine, zhao2021seeing, helber2019eurosat,bickel2021labeled,bai2016cloud, modava2017coastline}. Most deep learning networks on images leverage convolutional neural networks (CNNs) which self-learn filters for various tasks~\cite{stridedconvolution}. These images are then typically compressed using autoencoders (deep neural networks with narrow waist) to create signature optimized for a set of tasks~\cite{ryu2018residential,zhao2019accurate,yang2019deepattern}. These narrow outputs are used to communicate between the sensor and the receiver as they maximize information density for a set of tasks~\cite{jiang2019turbo}.  Unfortunately, if autoencoders are trained unaware of the communication channel, there is large information loss due to discretization, and bit errors ~\cite{yang2018bit}. There is of course a lot of work in joint source-channel coding for wireless image transmission\cite{guionnet2003soft,chandramouli1998adaptive,burlina1998error,debrunner2000error,bourtsoulatze2019deep} which deal with general uniform error distributions seen in typical wireless channels. However, operating LoRa (a technology that encodes data in frequency shifts) in 915 MHz ISM band (large amount of Doppler) leads to a non-uniform distribution of errors across bits as described in Sec.\ref{ssec:channel-aware}. In this paper, \name\ shows how, by making these autoencoders aware of this non-uniform distribution in the LoRa CubeSat context, we can design extremely small encodings robust to an adverse communication channel~(Sec.~\ref{sec:chaware}).

%% file: text/10-conclusion.tex
\section{Conclusion and Future Work}
This paper presents \name, a LoRa-based CubeSat communication system that allows delivery of quality of image information for wide-ranging tasks to commodity commercial ground stations. This is achieved through a LoRa channel-aware image encoding scheme that is informed by the structure of satellite images, the tasks performed on them, and the physical-layer channel characteristics of satellite signals. We present a detailed evaluation of \name\ by leveraging wireless channel measurements from from  recent CubeSat launched by our team. Our results demonstrate a 4.56~dB improvement of recovered image quality and the 1.38$\times$ improvement in accuracy of land-use classification. We plan to deploy \name\ end-to-end on a future CubeSat mission to estimate our system's ability for varied Earth sensing applications.

%% file: main_revision.bbl

\begin{thebibliography}{72}

%%% the Web address.

\ifx \showCODEN    \undefined \def \showCODEN     #1{\unskip}     \fi
\ifx \showDOI      \undefined \def \showDOI       #1{#1}\fi
\ifx \showISBNx    \undefined \def \showISBNx     #1{\unskip}     \fi
\ifx \showISBNxiii \undefined \def \showISBNxiii  #1{\unskip}     \fi
\ifx \showISSN     \undefined \def \showISSN      #1{\unskip}     \fi
\ifx \showLCCN     \undefined \def \showLCCN      #1{\unskip}     \fi
\ifx \shownote     \undefined \def \shownote      #1{#1}          \fi
\ifx \showarticletitle \undefined \def \showarticletitle #1{#1}   \fi
\ifx \showURL      \undefined \def \showURL       {\relax}        \fi
% invisible to TeX
\providecommand\bibfield[2]{#2}
\providecommand\bibinfo[2]{#2}
\providecommand\natexlab[1]{#1}
\providecommand\showeprint[2][]{arXiv:#2}

\bibitem[\protect\citeauthoryear{??}{vr3}{[n.d.]}]%
        {vr3x}
 \bibinfo{year}{[n.d.]}\natexlab{}.
\newblock \bibinfo{title}{Anonymized for Review}.
\newblock
\newblock


\bibitem[\protect\citeauthoryear{??}{tle}{[n.d.]}]%
        {tleorbit}
 \bibinfo{year}{[n.d.]}\natexlab{}.
\newblock \bibinfo{title}{{Assessment of TLE-based Orbit Determination and
  Prediction for Cubesats}}.
\newblock
  \bibinfo{howpublished}{\url{https://ntrs.nasa.gov/api/citations/20190004996/downloads/20190004996.pdf}}.
\newblock


\bibitem[\protect\citeauthoryear{??}{boe}{[n.d.]}]%
        {boeing}
 \bibinfo{year}{[n.d.]}\natexlab{}.
\newblock \bibinfo{title}{{Boeing Aerospace Satellites}}.
\newblock
  \bibinfo{howpublished}{\url{https://www.boeing.com/space/boeing-satellite-family/}}.
\newblock


\bibitem[\protect\citeauthoryear{??}{str}{[n.d.]}]%
        {stridedconvolution}
 \bibinfo{year}{[n.d.]}\natexlab{}.
\newblock \bibinfo{title}{Convolutional Neural Networks - The Math of
  Intelligence}.
\newblock
  \bibinfo{howpublished}{\url{https://www.youtube.com/watch?v=FTr3n7uBIuE}}.
\newblock


\bibitem[\protect\citeauthoryear{??}{zac}{[n.d.]}]%
        {zacpaper}
 \bibinfo{year}{[n.d.]}\natexlab{}.
\newblock \bibinfo{title}{CubeSat Compute, Storage and Control System Design
  Paper (Anonymized for Review)}.
\newblock
\newblock


\bibitem[\protect\citeauthoryear{??}{fos}{[n.d.]}]%
        {fossat1}
 \bibinfo{year}{[n.d.]}\natexlab{}.
\newblock \bibinfo{title}{{FossaSat-1, an Open Source Satellite for the
  Internet of Things}}.
\newblock
  \bibinfo{howpublished}{\url{https://www.hackster.io/news/fossasat-1-an-open-source-satellite-for-the-internet-of-things-7f31cab00ef5}}.
\newblock


\bibitem[\protect\citeauthoryear{??}{gos}{[n.d.]}]%
        {gossamer}
 \bibinfo{year}{[n.d.]}\natexlab{}.
\newblock \bibinfo{title}{{Gossamer Piccolomini (Gossamer 1)}}.
\newblock \bibinfo{howpublished}{\url{https://tinygs.com/satellite/Gossamer}}.
\newblock


\bibitem[\protect\citeauthoryear{??}{xor}{[n.d.]}]%
        {xorNN}
 \bibinfo{year}{[n.d.]}\natexlab{}.
\newblock \bibinfo{title}{{How neural networks solve the XOR problem}}.
\newblock
  \bibinfo{howpublished}{\url{https://towardsdatascience.com/how-neural-networks-solve-the-xor-problem-59763136bdd7}}.
\newblock


\bibitem[\protect\citeauthoryear{??}{lac}{[n.d.]}]%
        {lacunaspace}
 \bibinfo{year}{[n.d.]}\natexlab{}.
\newblock \bibinfo{title}{{Lacuna Space}}.
\newblock \bibinfo{howpublished}{\url{https://lacuna.space/}}.
\newblock


\bibitem[\protect\citeauthoryear{??}{lan}{[n.d.]}]%
        {landsat}
 \bibinfo{year}{[n.d.]}\natexlab{}.
\newblock \bibinfo{title}{{Landsat Science Missions}}.
\newblock \bibinfo{howpublished}{\url{https://landsat.gsfc.nasa.gov/}}.
\newblock


\bibitem[\protect\citeauthoryear{??}{noa}{[n.d.]}]%
        {noaa}
 \bibinfo{year}{[n.d.]}\natexlab{}.
\newblock \bibinfo{title}{{NOAA satellites}}.
\newblock \bibinfo{howpublished}{\url{https://www.noaa.gov/satellites}}.
\newblock


\bibitem[\protect\citeauthoryear{??}{pla}{[n.d.]}]%
        {planetlabs}
 \bibinfo{year}{[n.d.]}\natexlab{}.
\newblock \bibinfo{title}{{Planet Labs}}.
\newblock \bibinfo{howpublished}{\url{https://www.planet.com/}}.
\newblock


\bibitem[\protect\citeauthoryear{??}{sat}{[n.d.]a}]%
        {sattala}
 \bibinfo{year}{[n.d.]}\natexlab{a}.
\newblock \bibinfo{title}{{SATLLA-2B}}.
\newblock \bibinfo{howpublished}{\url{https://tinygs.com/satellite/SATLLA-2B}}.
\newblock


\bibitem[\protect\citeauthoryear{??}{sat}{[n.d.]b}]%
        {satnogsdb}
 \bibinfo{year}{[n.d.]}\natexlab{b}.
\newblock \bibinfo{title}{{SatNogs DB}}.
\newblock \bibinfo{howpublished}{\url{https://db.satnogs.org/}}.
\newblock


\bibitem[\protect\citeauthoryear{??}{sx1}{[n.d.]a}]%
        {sx1262calculator}
 \bibinfo{year}{[n.d.]}\natexlab{a}.
\newblock \bibinfo{title}{{SemTech SX1261 LoRa Calculator}}.
\newblock
  \bibinfo{howpublished}{\url{https://semtech.my.salesforce.com/sfc/p/\#E0000000JelG/a/2R000000Q2OT/GhbZe2lGVNO6sNDUlo6lcHVaKMQvcVCdaYfFeSjyitk}}.
\newblock


\bibitem[\protect\citeauthoryear{??}{sx1}{[n.d.]b}]%
        {sx1262}
 \bibinfo{year}{[n.d.]}\natexlab{b}.
\newblock \bibinfo{title}{{SemTech SX1262}}.
\newblock
  \bibinfo{howpublished}{\url{https://www.semtech.com/products/wireless-rf/lora-core/sx1262}}.
\newblock


\bibitem[\protect\citeauthoryear{??}{sen}{[n.d.]}]%
        {sentinel}
 \bibinfo{year}{[n.d.]}\natexlab{}.
\newblock \bibinfo{title}{{Sentinel Missions}}.
\newblock
  \bibinfo{howpublished}{\url{https://www.esa.int/Applications/Observing_the_Earth/Copernicus/The_Sentinel_missions}}.
\newblock


\bibitem[\protect\citeauthoryear{??}{swa}{[n.d.]}]%
        {swarm}
 \bibinfo{year}{[n.d.]}\natexlab{}.
\newblock \bibinfo{title}{{Swarm takes LoRa Sky High}}.
\newblock
  \bibinfo{howpublished}{\url{https://spectrum.ieee.org/swarm-takes-lora-skyhigh}}.
\newblock


\bibitem[\protect\citeauthoryear{??}{tin}{[n.d.]}]%
        {tinygs}
 \bibinfo{year}{[n.d.]}\natexlab{}.
\newblock \bibinfo{title}{{TinyGS, the Open Source Global Satellite Network}}.
\newblock \bibinfo{howpublished}{\url{https://tinygs.com/}}.
\newblock


\bibitem[\protect\citeauthoryear{??}{che}{[n.d.]}]%
        {cheaplaunch}
 \bibinfo{year}{[n.d.]}\natexlab{}.
\newblock \bibinfo{title}{{Why Space Launches are getting cheaper?}}
\newblock
  \bibinfo{howpublished}{\url{https://www.facebook.com/watch/?v=3421542817907662}}.
\newblock


\bibitem[\protect\citeauthoryear{??}{thi}{2017}]%
        {things-network}
 \bibinfo{year}{accessed Jan 10, 2017}\natexlab{}.
\newblock \bibinfo{title}{The Things Network}.
\newblock
\newblock
\urldef\tempurl%
\url{https://www.thethingsnetwork.org/}
\showURL{%
\tempurl}


\bibitem[\protect\citeauthoryear{Afisiadis, Cotting, Burg, and
  Balatsoukas-Stimming}{Afisiadis et~al\mbox{.}}{2019}]%
        {afisiadis2019error}
\bibfield{author}{\bibinfo{person}{Orion Afisiadis}, \bibinfo{person}{Matthieu
  Cotting}, \bibinfo{person}{Andreas Burg}, {and} \bibinfo{person}{Alexios
  Balatsoukas-Stimming}.} \bibinfo{year}{2019}\natexlab{}.
\newblock \showarticletitle{{On the error rate of the LoRa modulation with
  interference}}.
\newblock \bibinfo{journal}{\emph{IEEE Transactions on Wireless
  Communications}} \bibinfo{volume}{19}, \bibinfo{number}{2}
  (\bibinfo{year}{2019}), \bibinfo{pages}{1292--1304}.
\newblock


\bibitem[\protect\citeauthoryear{Agrawal, Barrington, Bromberg, Burge, Gazen,
  and Hickey}{Agrawal et~al\mbox{.}}{2019}]%
        {agrawal2019machine}
\bibfield{author}{\bibinfo{person}{Shreya Agrawal}, \bibinfo{person}{Luke
  Barrington}, \bibinfo{person}{Carla Bromberg}, \bibinfo{person}{John Burge},
  \bibinfo{person}{Cenk Gazen}, {and} \bibinfo{person}{Jason Hickey}.}
  \bibinfo{year}{2019}\natexlab{}.
\newblock \showarticletitle{Machine learning for precipitation nowcasting from
  radar images}.
\newblock \bibinfo{journal}{\emph{arXiv preprint arXiv:1912.12132}}
  (\bibinfo{year}{2019}).
\newblock


\bibitem[\protect\citeauthoryear{Ameloot, Moeneclaey, Van~Torre, and
  Rogier}{Ameloot et~al\mbox{.}}{2021}]%
        {ameloot2021characterizing}
\bibfield{author}{\bibinfo{person}{Thomas Ameloot}, \bibinfo{person}{Marc
  Moeneclaey}, \bibinfo{person}{Patrick Van~Torre}, {and}
  \bibinfo{person}{Hendrik Rogier}.} \bibinfo{year}{2021}\natexlab{}.
\newblock \showarticletitle{{Characterizing the Impact of Doppler Effects on
  Body-Centric LoRa Links with SDR}}.
\newblock \bibinfo{journal}{\emph{Sensors}} \bibinfo{volume}{21},
  \bibinfo{number}{12} (\bibinfo{year}{2021}), \bibinfo{pages}{4049}.
\newblock


\bibitem[\protect\citeauthoryear{Bai, Li, Sun, Chen, and Li}{Bai
  et~al\mbox{.}}{2016}]%
        {bai2016cloud}
\bibfield{author}{\bibinfo{person}{Ting Bai}, \bibinfo{person}{Deren Li},
  \bibinfo{person}{Kaimin Sun}, \bibinfo{person}{Yepei Chen}, {and}
  \bibinfo{person}{Wenzhuo Li}.} \bibinfo{year}{2016}\natexlab{}.
\newblock \showarticletitle{Cloud detection for high-resolution satellite
  imagery using machine learning and multi-feature fusion}.
\newblock \bibinfo{journal}{\emph{Remote Sensing}} \bibinfo{volume}{8},
  \bibinfo{number}{9} (\bibinfo{year}{2016}), \bibinfo{pages}{715}.
\newblock


\bibitem[\protect\citeauthoryear{Bickel, Mandrake, and Doran}{Bickel
  et~al\mbox{.}}{2021}]%
        {bickel2021labeled}
\bibfield{author}{\bibinfo{person}{Valentin Bickel}, \bibinfo{person}{Lukas
  Mandrake}, {and} \bibinfo{person}{Gary Doran}.}
  \bibinfo{year}{2021}\natexlab{}.
\newblock \showarticletitle{{A Labeled Image Dataset for Deep Learning-Driven
  Rockfall Detection on the Moon and Mars}}.
\newblock \bibinfo{journal}{\emph{Frontiers in Remote Sensing}}
  \bibinfo{volume}{2} (\bibinfo{year}{2021}), \bibinfo{pages}{640034}.
\newblock


\bibitem[\protect\citeauthoryear{Bohra, De~Meer, and Memon}{Bohra
  et~al\mbox{.}}{2009}]%
        {bohra2009analysing}
\bibfield{author}{\bibinfo{person}{Nafeesa Bohra}, \bibinfo{person}{Hermann
  De~Meer}, {and} \bibinfo{person}{Aftab~A Memon}.}
  \bibinfo{year}{2009}\natexlab{}.
\newblock \showarticletitle{{Analysing the Orbital Movement and Trajectory of
  LEO (Low Earth Orbit) Satellite Relative to Earth Rotation}}. In
  \bibinfo{booktitle}{\emph{International Conference on Personal Satellite
  Services}}. Springer, \bibinfo{pages}{1--11}.
\newblock


\bibitem[\protect\citeauthoryear{Bourtsoulatze, Kurka, and
  G{\"u}nd{\"u}z}{Bourtsoulatze et~al\mbox{.}}{2019}]%
        {bourtsoulatze2019deep}
\bibfield{author}{\bibinfo{person}{Eirina Bourtsoulatze},
  \bibinfo{person}{David~Burth Kurka}, {and} \bibinfo{person}{Deniz
  G{\"u}nd{\"u}z}.} \bibinfo{year}{2019}\natexlab{}.
\newblock \showarticletitle{Deep joint source-channel coding for wireless image
  transmission}.
\newblock \bibinfo{journal}{\emph{IEEE Transactions on Cognitive Communications
  and Networking}} \bibinfo{volume}{5}, \bibinfo{number}{3}
  (\bibinfo{year}{2019}), \bibinfo{pages}{567--579}.
\newblock


\bibitem[\protect\citeauthoryear{Burlina and Alajaji}{Burlina and
  Alajaji}{1998}]%
        {burlina1998error}
\bibfield{author}{\bibinfo{person}{Philippe Burlina} {and}
  \bibinfo{person}{Fady Alajaji}.} \bibinfo{year}{1998}\natexlab{}.
\newblock \showarticletitle{An error resilient scheme for image transmission
  over noisy channels with memory}.
\newblock \bibinfo{journal}{\emph{IEEE Transactions on Image Processing}}
  \bibinfo{volume}{7}, \bibinfo{number}{4} (\bibinfo{year}{1998}),
  \bibinfo{pages}{593--600}.
\newblock


\bibitem[\protect\citeauthoryear{Chakrabarti}{Chakrabarti}{[n.d.]}]%
        {alotofsats}
\bibfield{author}{\bibinfo{person}{Supriya Chakrabarti}.}
  \bibinfo{year}{[n.d.]}\natexlab{}.
\newblock \bibinfo{title}{{How many satellites are orbiting Earth?}}
\newblock
  \bibinfo{howpublished}{\url{https://astronomy.com/news/2021/09/how-many-satellites-are-orbiting-earth}}.
\newblock


\bibitem[\protect\citeauthoryear{Chandramouli, Ranganathan, and
  Ramadoss}{Chandramouli et~al\mbox{.}}{1998}]%
        {chandramouli1998adaptive}
\bibfield{author}{\bibinfo{person}{Ramamurti Chandramouli}, \bibinfo{person}{N
  Ranganathan}, {and} \bibinfo{person}{Shivaraman~J Ramadoss}.}
  \bibinfo{year}{1998}\natexlab{}.
\newblock \showarticletitle{Adaptive quantization and fast error-resilient
  entropy coding for image transmission}.
\newblock \bibinfo{journal}{\emph{IEEE transactions on circuits and systems for
  video technology}} \bibinfo{volume}{8}, \bibinfo{number}{4}
  (\bibinfo{year}{1998}), \bibinfo{pages}{411--421}.
\newblock


\bibitem[\protect\citeauthoryear{Chen, Dong, and Lv}{Chen
  et~al\mbox{.}}{2021}]%
        {chen2021lofi}
\bibfield{author}{\bibinfo{person}{Gonglong Chen}, \bibinfo{person}{Wei Dong},
  {and} \bibinfo{person}{Jiamei Lv}.} \bibinfo{year}{2021}\natexlab{}.
\newblock \showarticletitle{{LoFi: Enabling 2.4 GHz LoRa and WiFi Coexistence
  by Detecting Extremely Weak Signals}}. In \bibinfo{booktitle}{\emph{IEEE
  INFOCOM 2021-IEEE Conference on Computer Communications}}. IEEE,
  \bibinfo{pages}{1--10}.
\newblock


\bibitem[\protect\citeauthoryear{Commission}{Commission}{2021a}]%
        {fcc}
\bibfield{author}{\bibinfo{person}{Federal~Communications Commission}.}
  \bibinfo{year}{2021}\natexlab{a}.
\newblock \bibinfo{title}{{Federal Communications Commission}}.
\newblock \bibinfo{howpublished}{\url{https://www.fcc.gov/}}.
\newblock


\bibitem[\protect\citeauthoryear{Commission}{Commission}{2021b}]%
        {fccspectrum}
\bibfield{author}{\bibinfo{person}{Federal~Communications Commission}.}
  \bibinfo{year}{2021}\natexlab{b}.
\newblock \bibinfo{title}{{Radio Spectrum Allocation}}.
\newblock
  \bibinfo{howpublished}{\url{https://www.fcc.gov/engineering-technology/policy-and-rules-division/general/radio-spectrum-allocation}}.
\newblock


\bibitem[\protect\citeauthoryear{Debrunner, DeBrunner, Wang, and
  Radhakrishnan}{Debrunner et~al\mbox{.}}{2000}]%
        {debrunner2000error}
\bibfield{author}{\bibinfo{person}{Victor Debrunner}, \bibinfo{person}{Linda
  DeBrunner}, \bibinfo{person}{Longji Wang}, {and} \bibinfo{person}{Sridhar
  Radhakrishnan}.} \bibinfo{year}{2000}\natexlab{}.
\newblock \showarticletitle{Error control and concealment for image
  transmission}.
\newblock \bibinfo{journal}{\emph{IEEE Communications Surveys \& Tutorials}}
  \bibinfo{volume}{3}, \bibinfo{number}{1} (\bibinfo{year}{2000}),
  \bibinfo{pages}{2--9}.
\newblock


\bibitem[\protect\citeauthoryear{Dongare, Narayanan, Gadre, Luong, Balanuta,
  Kumar, Iannucci, and Rowe}{Dongare et~al\mbox{.}}{2018}]%
        {dongare2018charm}
\bibfield{author}{\bibinfo{person}{Adwait Dongare}, \bibinfo{person}{Revathy
  Narayanan}, \bibinfo{person}{Akshay Gadre}, \bibinfo{person}{Anh Luong},
  \bibinfo{person}{Artur Balanuta}, \bibinfo{person}{Swarun Kumar},
  \bibinfo{person}{Bob Iannucci}, {and} \bibinfo{person}{Anthony Rowe}.}
  \bibinfo{year}{2018}\natexlab{}.
\newblock \showarticletitle{Charm: exploiting geographical diversity through
  coherent combining in low-power wide-area networks}. In
  \bibinfo{booktitle}{\emph{2018 17th ACM/IEEE International Conference on
  Information Processing in Sensor Networks (IPSN)}}. IEEE,
  \bibinfo{pages}{60--71}.
\newblock


\bibitem[\protect\citeauthoryear{Doroshkin, Zadorozhny, Kus, Prokopyev, and
  Prokopyev}{Doroshkin et~al\mbox{.}}{2019}]%
        {doroshkin2019experimental}
\bibfield{author}{\bibinfo{person}{Alexander~A Doroshkin},
  \bibinfo{person}{Alexander~M Zadorozhny}, \bibinfo{person}{Oleg~N Kus},
  \bibinfo{person}{Vitaliy~Yu Prokopyev}, {and} \bibinfo{person}{Yuri~M
  Prokopyev}.} \bibinfo{year}{2019}\natexlab{}.
\newblock \showarticletitle{Experimental study of LoRa modulation immunity to
  Doppler effect in CubeSat radio communications}.
\newblock \bibinfo{journal}{\emph{IEEE Access}}  \bibinfo{volume}{7}
  (\bibinfo{year}{2019}), \bibinfo{pages}{75721--75731}.
\newblock


\bibitem[\protect\citeauthoryear{Fernandez, Ruiz-De-Azua, Calveras, and
  Camps}{Fernandez et~al\mbox{.}}{2020}]%
        {fernandez2020assessing}
\bibfield{author}{\bibinfo{person}{Lara Fernandez}, \bibinfo{person}{Joan~Adria
  Ruiz-De-Azua}, \bibinfo{person}{Anna Calveras}, {and}
  \bibinfo{person}{Adriano Camps}.} \bibinfo{year}{2020}\natexlab{}.
\newblock \showarticletitle{Assessing LoRa for satellite-to-earth
  communications considering the impact of ionospheric scintillation}.
\newblock \bibinfo{journal}{\emph{IEEE access}}  \bibinfo{volume}{8}
  (\bibinfo{year}{2020}), \bibinfo{pages}{165570--165582}.
\newblock


\bibitem[\protect\citeauthoryear{Gadre, Narayanan, Luong, Rowe, Iannucci, and
  Kumar}{Gadre et~al\mbox{.}}{2020a}]%
        {gadre2020frequency}
\bibfield{author}{\bibinfo{person}{Akshay Gadre}, \bibinfo{person}{Revathy
  Narayanan}, \bibinfo{person}{Anh Luong}, \bibinfo{person}{Anthony Rowe},
  \bibinfo{person}{Bob Iannucci}, {and} \bibinfo{person}{Swarun Kumar}.}
  \bibinfo{year}{2020}\natexlab{a}.
\newblock \showarticletitle{Frequency configuration for low-power wide-area
  networks in a heartbeat}. In \bibinfo{booktitle}{\emph{17th $\{$USENIX$\}$
  Symposium on Networked Systems Design and Implementation ($\{$NSDI$\}$ 20)}}.
  \bibinfo{pages}{339--352}.
\newblock


\bibitem[\protect\citeauthoryear{Gadre, Yi, Rowe, Iannucci, and Kumar}{Gadre
  et~al\mbox{.}}{2020b}]%
        {gadre2020quick}
\bibfield{author}{\bibinfo{person}{Akshay Gadre}, \bibinfo{person}{Fan Yi},
  \bibinfo{person}{Anthony Rowe}, \bibinfo{person}{Bob Iannucci}, {and}
  \bibinfo{person}{Swarun Kumar}.} \bibinfo{year}{2020}\natexlab{b}.
\newblock \showarticletitle{{Quick (and Dirty) Aggregate Queries on Low-Power
  WANs}}. In \bibinfo{booktitle}{\emph{2020 19th ACM/IEEE International
  Conference on Information Processing in Sensor Networks (IPSN)}}. IEEE,
  \bibinfo{pages}{277--288}.
\newblock


\bibitem[\protect\citeauthoryear{Gamage, Liando, Gu, Tan, and Li}{Gamage
  et~al\mbox{.}}{2020}]%
        {gamage2020lmac}
\bibfield{author}{\bibinfo{person}{Amalinda Gamage},
  \bibinfo{person}{Jansen~Christian Liando}, \bibinfo{person}{Chaojie Gu},
  \bibinfo{person}{Rui Tan}, {and} \bibinfo{person}{Mo Li}.}
  \bibinfo{year}{2020}\natexlab{}.
\newblock \showarticletitle{{LMAC: Efficient carrier-sense multiple access for
  lora}}. In \bibinfo{booktitle}{\emph{Proceedings of the 26th Annual
  International Conference on Mobile Computing and Networking}}.
  \bibinfo{pages}{1--13}.
\newblock


\bibitem[\protect\citeauthoryear{Guier and Weiffenbach}{Guier and
  Weiffenbach}{1959}]%
        {guier1959doppler}
\bibfield{author}{\bibinfo{person}{William~H Guier} {and}
  \bibinfo{person}{George~C Weiffenbach}.} \bibinfo{year}{1959}\natexlab{}.
\newblock \bibinfo{booktitle}{\emph{The Doppler determination of orbits}}.
\newblock \bibinfo{type}{{T}echnical {R}eport}. \bibinfo{institution}{JOHNS
  HOPKINS UNIV LAUREL MD APPLIED PHYSICS LAB}.
\newblock


\bibitem[\protect\citeauthoryear{Guier and Weiffenbach}{Guier and
  Weiffenbach}{1998}]%
        {guier1998genesis}
\bibfield{author}{\bibinfo{person}{William~H Guier} {and}
  \bibinfo{person}{George~C Weiffenbach}.} \bibinfo{year}{1998}\natexlab{}.
\newblock \showarticletitle{Genesis of satellite navigation}.
\newblock \bibinfo{journal}{\emph{Johns Hopkins APL technical digest}}
  \bibinfo{volume}{19}, \bibinfo{number}{1} (\bibinfo{year}{1998}),
  \bibinfo{pages}{15}.
\newblock


\bibitem[\protect\citeauthoryear{Guionnet and Guillemot}{Guionnet and
  Guillemot}{2003}]%
        {guionnet2003soft}
\bibfield{author}{\bibinfo{person}{Thomas Guionnet} {and}
  \bibinfo{person}{Christine Guillemot}.} \bibinfo{year}{2003}\natexlab{}.
\newblock \showarticletitle{Soft decoding and synchronization of arithmetic
  codes: Application to image transmission over noisy channels}.
\newblock \bibinfo{journal}{\emph{IEEE transactions on Image Processing}}
  \bibinfo{volume}{12}, \bibinfo{number}{12} (\bibinfo{year}{2003}),
  \bibinfo{pages}{1599--1609}.
\newblock


\bibitem[\protect\citeauthoryear{Helber, Bischke, Dengel, and Borth}{Helber
  et~al\mbox{.}}{2019}]%
        {helber2019eurosat}
\bibfield{author}{\bibinfo{person}{Patrick Helber}, \bibinfo{person}{Benjamin
  Bischke}, \bibinfo{person}{Andreas Dengel}, {and} \bibinfo{person}{Damian
  Borth}.} \bibinfo{year}{2019}\natexlab{}.
\newblock \showarticletitle{Eurosat: A novel dataset and deep learning
  benchmark for land use and land cover classification}.
\newblock \bibinfo{journal}{\emph{IEEE Journal of Selected Topics in Applied
  Earth Observations and Remote Sensing}} \bibinfo{volume}{12},
  \bibinfo{number}{7} (\bibinfo{year}{2019}), \bibinfo{pages}{2217--2226}.
\newblock


\bibitem[\protect\citeauthoryear{Hessar, Najafi, and Gollakota}{Hessar
  et~al\mbox{.}}{2019}]%
        {hessar2019netscatter}
\bibfield{author}{\bibinfo{person}{Mehrdad Hessar}, \bibinfo{person}{Ali
  Najafi}, {and} \bibinfo{person}{Shyamnath Gollakota}.}
  \bibinfo{year}{2019}\natexlab{}.
\newblock \showarticletitle{{Netscatter: Enabling large-scale backscatter
  networks}}. In \bibinfo{booktitle}{\emph{16th $\{$USENIX$\}$ Symposium on
  Networked Systems Design and Implementation ($\{$NSDI$\}$ 19)}}.
  \bibinfo{pages}{271--284}.
\newblock


\bibitem[\protect\citeauthoryear{Jiang, Kim, Asnani, Kannan, Oh, and
  Viswanath}{Jiang et~al\mbox{.}}{2019}]%
        {jiang2019turbo}
\bibfield{author}{\bibinfo{person}{Yihan Jiang}, \bibinfo{person}{Hyeji Kim},
  \bibinfo{person}{Himanshu Asnani}, \bibinfo{person}{Sreeram Kannan},
  \bibinfo{person}{Sewoong Oh}, {and} \bibinfo{person}{Pramod Viswanath}.}
  \bibinfo{year}{2019}\natexlab{}.
\newblock \showarticletitle{{Turbo autoencoder: Deep learning based channel
  codes for point-to-point communication channels}}.
\newblock \bibinfo{journal}{\emph{Advances in neural information processing
  systems}}  \bibinfo{volume}{32} (\bibinfo{year}{2019}),
  \bibinfo{pages}{2758--2768}.
\newblock


\bibitem[\protect\citeauthoryear{Khalife, Neinavaie, and Kassas}{Khalife
  et~al\mbox{.}}{2020}]%
        {khalife2020blind}
\bibfield{author}{\bibinfo{person}{Joe Khalife}, \bibinfo{person}{Mohammad
  Neinavaie}, {and} \bibinfo{person}{Zaher~M Kassas}.}
  \bibinfo{year}{2020}\natexlab{}.
\newblock \showarticletitle{{Blind Doppler estimation from LEO satellite
  signals: A case study with real 5G signals}}. In
  \bibinfo{booktitle}{\emph{Proceedings of the 33rd International Technical
  Meeting of the Satellite Division of The Institute of Navigation (ION GNSS+
  2020)}}. \bibinfo{pages}{3046--3054}.
\newblock


\bibitem[\protect\citeauthoryear{Klofas, Anderson, and Leveque}{Klofas
  et~al\mbox{.}}{2008}]%
        {klofas2008survey}
\bibfield{author}{\bibinfo{person}{Bryan Klofas}, \bibinfo{person}{Jason
  Anderson}, {and} \bibinfo{person}{Kyle Leveque}.}
  \bibinfo{year}{2008}\natexlab{}.
\newblock \showarticletitle{A survey of cubesat communication systems}. In
  \bibinfo{booktitle}{\emph{5th Annual CubeSat Developers’ Workshop}}.
  \bibinfo{pages}{1--36}.
\newblock


\bibitem[\protect\citeauthoryear{Lapapan, Pa-in, Chotimanon, and
  Chodkaveekityada}{Lapapan et~al\mbox{.}}{2021}]%
        {lapapan2021lora}
\bibfield{author}{\bibinfo{person}{Intanon Lapapan},
  \bibinfo{person}{Wetchaphat Pa-in}, \bibinfo{person}{Vachiravit Chotimanon},
  {and} \bibinfo{person}{Peeramed Chodkaveekityada}.}
  \bibinfo{year}{2021}\natexlab{}.
\newblock \showarticletitle{{LoRa Multi-Channel Access to Doppler effect in
  CubeSat Radio Communication}}.
\newblock  (\bibinfo{year}{2021}).
\newblock


\bibitem[\protect\citeauthoryear{Modava and Akbarizadeh}{Modava and
  Akbarizadeh}{2017}]%
        {modava2017coastline}
\bibfield{author}{\bibinfo{person}{Mohammad Modava} {and}
  \bibinfo{person}{Gholamreza Akbarizadeh}.} \bibinfo{year}{2017}\natexlab{}.
\newblock \showarticletitle{{Coastline extraction from SAR images using spatial
  fuzzy clustering and the active contour method}}.
\newblock \bibinfo{journal}{\emph{International journal of remote sensing}}
  \bibinfo{volume}{38}, \bibinfo{number}{2} (\bibinfo{year}{2017}),
  \bibinfo{pages}{355--370}.
\newblock


\bibitem[\protect\citeauthoryear{Nardello, Desai, Brunelli, and Lucia}{Nardello
  et~al\mbox{.}}{2019}]%
        {nardello2019camaroptera}
\bibfield{author}{\bibinfo{person}{Matteo Nardello}, \bibinfo{person}{Harsh
  Desai}, \bibinfo{person}{Davide Brunelli}, {and} \bibinfo{person}{Brandon
  Lucia}.} \bibinfo{year}{2019}\natexlab{}.
\newblock \showarticletitle{{Camaroptera: A batteryless long-range remote
  visual sensing system}}. In \bibinfo{booktitle}{\emph{Proceedings of the 7th
  International Workshop on Energy Harvesting \& Energy-Neutral Sensing
  Systems}}. \bibinfo{pages}{8--14}.
\newblock


\bibitem[\protect\citeauthoryear{{National Telecommunications and Information
  Administration}}{{National Telecommunications and Information
  Administration}}{2021a}]%
        {ntiafcc}
\bibfield{author}{\bibinfo{person}{{National Telecommunications and Information
  Administration}}.} \bibinfo{year}{2021}\natexlab{a}.
\newblock \bibinfo{title}{{FCC, NTIA Name Staff Representatives to Advisory
  Committees to Further Technical Collaboration}}.
\newblock
  \bibinfo{howpublished}{\url{https://www.ntia.doc.gov/press-release/2022/fcc-ntia-name-staff-representatives-advisory-committees-further-technical}}.
\newblock


\bibitem[\protect\citeauthoryear{{National Telecommunications and Information
  Administration}}{{National Telecommunications and Information
  Administration}}{2021b}]%
        {ntia}
\bibfield{author}{\bibinfo{person}{{National Telecommunications and Information
  Administration}}.} \bibinfo{year}{2021}\natexlab{b}.
\newblock \bibinfo{title}{{National Telecommunications and Information
  Administration}}.
\newblock \bibinfo{howpublished}{\url{https://www.ntia.doc.gov/}}.
\newblock


\bibitem[\protect\citeauthoryear{Osborne and Xie}{Osborne and Xie}{1999}]%
        {osborne1999propagation}
\bibfield{author}{\bibinfo{person}{William~P Osborne} {and}
  \bibinfo{person}{Yongjun Xie}.} \bibinfo{year}{1999}\natexlab{}.
\newblock \showarticletitle{{Propagation characterization of LEO/MEO satellite
  systems at 900-2100 MHz}}. In \bibinfo{booktitle}{\emph{1999 IEEE Emerging
  Technologies Symposium. Wireless Communications and Systems (IEEE Cat. No.
  99EX297)}}. IEEE, \bibinfo{pages}{21--1}.
\newblock


\bibitem[\protect\citeauthoryear{Palattella and Accettura}{Palattella and
  Accettura}{2018}]%
        {palattella2018enabling}
\bibfield{author}{\bibinfo{person}{Maria~Rita Palattella} {and}
  \bibinfo{person}{Nicola Accettura}.} \bibinfo{year}{2018}\natexlab{}.
\newblock \showarticletitle{Enabling internet of everything everywhere: LPWAN
  with satellite backhaul}. In \bibinfo{booktitle}{\emph{2018 Global
  Information Infrastructure and Networking Symposium (GIIS)}}. IEEE,
  \bibinfo{pages}{1--5}.
\newblock


\bibitem[\protect\citeauthoryear{Ryu, Choi, Lee, Kim, and Wong}{Ryu
  et~al\mbox{.}}{2018}]%
        {ryu2018residential}
\bibfield{author}{\bibinfo{person}{Seunghyoung Ryu}, \bibinfo{person}{Hyungeun
  Choi}, \bibinfo{person}{Hyoseop Lee}, \bibinfo{person}{Hongseok Kim}, {and}
  \bibinfo{person}{Vincent~WS Wong}.} \bibinfo{year}{2018}\natexlab{}.
\newblock \showarticletitle{{Residential load profile clustering via deep
  convolutional autoencoder}}. In \bibinfo{booktitle}{\emph{2018 IEEE
  international conference on communications, control, and computing
  technologies for smart grids (SmartGridComm)}}. IEEE, \bibinfo{pages}{1--6}.
\newblock


\bibitem[\protect\citeauthoryear{Shahid, Philipose, Chintalapudi, Banerjee, and
  Krishnaswamy}{Shahid et~al\mbox{.}}{2021}]%
        {shahid2021concurrent}
\bibfield{author}{\bibinfo{person}{Muhammad~Osama Shahid},
  \bibinfo{person}{Millan Philipose}, \bibinfo{person}{Krishna Chintalapudi},
  \bibinfo{person}{Suman Banerjee}, {and} \bibinfo{person}{Bhuvana
  Krishnaswamy}.} \bibinfo{year}{2021}\natexlab{}.
\newblock \showarticletitle{{Concurrent interference cancellation: decoding
  multi-packet collisions in LoRa}}. In \bibinfo{booktitle}{\emph{Proceedings
  of the 2021 ACM SIGCOMM 2021 Conference}}. \bibinfo{pages}{503--515}.
\newblock


\bibitem[\protect\citeauthoryear{Shen, Huang, Song, Vincent, and Togneri}{Shen
  et~al\mbox{.}}{2019}]%
        {shen20193}
\bibfield{author}{\bibinfo{person}{Xi Shen}, \bibinfo{person}{Defeng~David
  Huang}, \bibinfo{person}{Boming Song}, \bibinfo{person}{Claire Vincent},
  {and} \bibinfo{person}{Roberto Togneri}.} \bibinfo{year}{2019}\natexlab{}.
\newblock \showarticletitle{{3-D tomographic reconstruction of rain field using
  microwave signals from LEO satellites: Principle and simulation results}}.
\newblock \bibinfo{journal}{\emph{IEEE Transactions on Geoscience and Remote
  Sensing}} \bibinfo{volume}{57}, \bibinfo{number}{8} (\bibinfo{year}{2019}),
  \bibinfo{pages}{5434--5446}.
\newblock


\bibitem[\protect\citeauthoryear{Singh, Prabhakara, Zhang, Ya{\u{g}}an, and
  Kumar}{Singh et~al\mbox{.}}{2021}]%
        {singh2021community}
\bibfield{author}{\bibinfo{person}{Vaibhav Singh}, \bibinfo{person}{Akarsh
  Prabhakara}, \bibinfo{person}{Diana Zhang}, \bibinfo{person}{Osman
  Ya{\u{g}}an}, {and} \bibinfo{person}{Swarun Kumar}.}
  \bibinfo{year}{2021}\natexlab{}.
\newblock \showarticletitle{{A community-driven approach to democratize access
  to satellite ground stations}}. In \bibinfo{booktitle}{\emph{Proceedings of
  the 27th Annual International Conference on Mobile Computing and
  Networking}}. \bibinfo{pages}{1--14}.
\newblock


\bibitem[\protect\citeauthoryear{Singh, Ya{\u{g}}an, and Kumar}{Singh
  et~al\mbox{.}}{2022}]%
        {singh2022selfiestick}
\bibfield{author}{\bibinfo{person}{Vaibhav Singh}, \bibinfo{person}{Osman
  Ya{\u{g}}an}, {and} \bibinfo{person}{Swarun Kumar}.}
  \bibinfo{year}{2022}\natexlab{}.
\newblock \showarticletitle{{SelfieStick: Towards Earth Imaging from a Low-Cost
  Ground Module Using LEO Satellites}}. In \bibinfo{booktitle}{\emph{ACM/IEEE
  International Conference on Information Processing in Sensor Networks
  (IPSN)}}.
\newblock


\bibitem[\protect\citeauthoryear{Sun, Yin, Chen, Wang, Zhang, and He}{Sun
  et~al\mbox{.}}{2021}]%
        {sun2021partial}
\bibfield{author}{\bibinfo{person}{Kai Sun}, \bibinfo{person}{Zhimeng Yin},
  \bibinfo{person}{Weiwei Chen}, \bibinfo{person}{Shuai Wang},
  \bibinfo{person}{Zeyu Zhang}, {and} \bibinfo{person}{Tian He}.}
  \bibinfo{year}{2021}\natexlab{}.
\newblock \showarticletitle{{Partial Symbol Recovery for Interference
  Resilience in Low-Power Wide Area Networks}}.
\newblock \bibinfo{journal}{\emph{arXiv preprint arXiv:2109.03488}}
  (\bibinfo{year}{2021}).
\newblock


\bibitem[\protect\citeauthoryear{{The Things Network}}{{The Things
  Network}}{2020}]%
        {lorarecord}
\bibfield{author}{\bibinfo{person}{{The Things Network}}.}
  \bibinfo{year}{2020}\natexlab{}.
\newblock \bibinfo{title}{{LoRa World Record Broken: 832km/517mi using 25mW}}.
\newblock
  \bibinfo{howpublished}{\url{https://www.thethingsnetwork.org/article/lorawan-world-record-broken-twice-in-single-experiment-1}}.
\newblock


\bibitem[\protect\citeauthoryear{Vasisht and Chandra}{Vasisht and
  Chandra}{2020}]%
        {vasisht2020distributed}
\bibfield{author}{\bibinfo{person}{Deepak Vasisht} {and}
  \bibinfo{person}{Ranveer Chandra}.} \bibinfo{year}{2020}\natexlab{}.
\newblock \showarticletitle{{A Distributed and Hybrid Ground Station Network
  for Low Earth Orbit Satellites}}. In \bibinfo{booktitle}{\emph{Proceedings of
  the 19th ACM Workshop on Hot Topics in Networks}}. \bibinfo{pages}{190--196}.
\newblock


\bibitem[\protect\citeauthoryear{Vasisht, Shenoy, and Chandra}{Vasisht
  et~al\mbox{.}}{2021}]%
        {vasisht2021l2d2}
\bibfield{author}{\bibinfo{person}{Deepak Vasisht}, \bibinfo{person}{Jayanth
  Shenoy}, {and} \bibinfo{person}{Ranveer Chandra}.}
  \bibinfo{year}{2021}\natexlab{}.
\newblock \showarticletitle{{L2D2: low latency distributed downlink for LEO
  satellites}}. In \bibinfo{booktitle}{\emph{Proceedings of the 2021 ACM
  SIGCOMM 2021 Conference}}. \bibinfo{pages}{151--164}.
\newblock


\bibitem[\protect\citeauthoryear{Voigt, Bor, Roedig, and Alonso}{Voigt
  et~al\mbox{.}}{2016}]%
        {voigt2016mitigating}
\bibfield{author}{\bibinfo{person}{Thiemo Voigt}, \bibinfo{person}{Martin Bor},
  \bibinfo{person}{Utz Roedig}, {and} \bibinfo{person}{Juan Alonso}.}
  \bibinfo{year}{2016}\natexlab{}.
\newblock \showarticletitle{{Mitigating inter-network interference in LoRa
  networks}}.
\newblock \bibinfo{journal}{\emph{arXiv preprint arXiv:1611.00688}}
  (\bibinfo{year}{2016}).
\newblock


\bibitem[\protect\citeauthoryear{Yang, Pathak, Gennari, Lai, and Yu}{Yang
  et~al\mbox{.}}{2019}]%
        {yang2019deepattern}
\bibfield{author}{\bibinfo{person}{Haoyu Yang}, \bibinfo{person}{Piyush
  Pathak}, \bibinfo{person}{Frank Gennari}, \bibinfo{person}{Ya-Chieh Lai},
  {and} \bibinfo{person}{Bei Yu}.} \bibinfo{year}{2019}\natexlab{}.
\newblock \showarticletitle{{DeePattern: Layout pattern generation with
  transforming convolutional auto-encoder}}. In
  \bibinfo{booktitle}{\emph{Proceedings of the 56th Annual Design Automation
  Conference 2019}}. \bibinfo{pages}{1--6}.
\newblock


\bibitem[\protect\citeauthoryear{Yang, Bankman, Moons, Verhelst, and
  Murmann}{Yang et~al\mbox{.}}{2018}]%
        {yang2018bit}
\bibfield{author}{\bibinfo{person}{Lita Yang}, \bibinfo{person}{Daniel
  Bankman}, \bibinfo{person}{Bert Moons}, \bibinfo{person}{Marian Verhelst},
  {and} \bibinfo{person}{Boris Murmann}.} \bibinfo{year}{2018}\natexlab{}.
\newblock \showarticletitle{{Bit error tolerance of a CIFAR-10 binarized
  convolutional neural network processor}}. In \bibinfo{booktitle}{\emph{2018
  IEEE International Symposium on Circuits and Systems (ISCAS)}}. IEEE,
  \bibinfo{pages}{1--5}.
\newblock


\bibitem[\protect\citeauthoryear{Yang, Zhang, Gadre, Liu, Kumar, and
  Sekar}{Yang et~al\mbox{.}}{2020}]%
        {yang2020joltik}
\bibfield{author}{\bibinfo{person}{Mingran Yang}, \bibinfo{person}{Junbo
  Zhang}, \bibinfo{person}{Akshay Gadre}, \bibinfo{person}{Zaoxing Liu},
  \bibinfo{person}{Swarun Kumar}, {and} \bibinfo{person}{Vyas Sekar}.}
  \bibinfo{year}{2020}\natexlab{}.
\newblock \showarticletitle{Joltik: enabling energy-efficient ``future-proof"
  analytics on low-power wide-area networks}. In
  \bibinfo{booktitle}{\emph{Proceedings of the 26th Annual International
  Conference on Mobile Computing and Networking}}. \bibinfo{pages}{1--14}.
\newblock


\bibitem[\protect\citeauthoryear{Yeh, Wu, Ko, and Wang}{Yeh
  et~al\mbox{.}}{2017}]%
        {yeh2017learning}
\bibfield{author}{\bibinfo{person}{Chih-Kuan Yeh}, \bibinfo{person}{Wei-Chieh
  Wu}, \bibinfo{person}{Wei-Jen Ko}, {and} \bibinfo{person}{Yu-Chiang~Frank
  Wang}.} \bibinfo{year}{2017}\natexlab{}.
\newblock \showarticletitle{Learning deep latent space for multi-label
  classification}. In \bibinfo{booktitle}{\emph{Thirty-first AAAI conference on
  artificial intelligence}}.
\newblock


\bibitem[\protect\citeauthoryear{Zhao, Huang, Li, Ding, Zhao, and Han}{Zhao
  et~al\mbox{.}}{2019}]%
        {zhao2019accurate}
\bibfield{author}{\bibinfo{person}{Lingjun Zhao}, \bibinfo{person}{Huakun
  Huang}, \bibinfo{person}{Xiang Li}, \bibinfo{person}{Shuxue Ding},
  \bibinfo{person}{Haoli Zhao}, {and} \bibinfo{person}{Zhaoyang Han}.}
  \bibinfo{year}{2019}\natexlab{}.
\newblock \showarticletitle{An accurate and robust approach of device-free
  localization with convolutional autoencoder}.
\newblock \bibinfo{journal}{\emph{IEEE Internet of Things Journal}}
  \bibinfo{volume}{6}, \bibinfo{number}{3} (\bibinfo{year}{2019}),
  \bibinfo{pages}{5825--5840}.
\newblock


\bibitem[\protect\citeauthoryear{Zhao, Olsen, and Chandra}{Zhao
  et~al\mbox{.}}{2021}]%
        {zhao2021seeing}
\bibfield{author}{\bibinfo{person}{Mingmin Zhao}, \bibinfo{person}{Peder~A
  Olsen}, {and} \bibinfo{person}{Ranveer Chandra}.}
  \bibinfo{year}{2021}\natexlab{}.
\newblock \showarticletitle{{Seeing Through Clouds in Satellite Images}}.
\newblock \bibinfo{journal}{\emph{arXiv preprint:2106.08408}}
  (\bibinfo{year}{2021}).
\newblock


\end{thebibliography}
